\documentclass[journal]{IEEEtran}
\usepackage{amsmath,amsfonts}
\usepackage{algorithmic}
\usepackage{algorithm}
\usepackage{array}
\usepackage[caption=false,font=footnotesize,labelfont=rm,textfont=rm]{subfig}
\usepackage{textcomp}
\usepackage{stfloats}
\usepackage{url}
\usepackage{verbatim}
\usepackage{graphicx}
\usepackage{cite}
\usepackage{color}
\usepackage{booktabs}
\usepackage{multirow}
\usepackage{makecell}
\usepackage{multicol}

\usepackage[switch]{lineno}

\begin{document}
\title{Motion Vector-Domain Video Steganalysis Exploiting Skipped Macroblocks}

\author{Jun Li, Minqing Zhang, Ke Niu, Yingnan Zhang, Xiaoyuan Yang.
\thanks{This work was supported by the National Natural Science Foundation of China (Grant No.62272478, No.62202496, No.62102450). (Corresponding author: Minqing Zhang.)
	
The authors are with the College of Cryptography Engineering in Engineering University of the Chinese People’s Armed Police Force, Xi’an, 710086, China; (e-mail: lijun9250lj@163.com, api\_zmq@126.com, niuke@163.com, zyn583@163.com, yxyangyxyang@163.com).}
}

\markboth{ }%
{Shell \MakeLowercase{\textit{Jun Li et al.}}: Motion Vector-Domain Video Steganalysis Exploiting Skipped Macroblocks}


\maketitle

\pagestyle{empty}
\thispagestyle{empty}

\begin{abstract}
Video steganography has the potential to be used to convey illegal information, and video steganalysis is a vital tool to detect the presence of this illicit act. Currently, all the motion vector (MV)-based video steganalysis algorithms extract feature sets directly on the MVs, but ignoring the steganograhic operation may perturb the statistics distribution of other video encoding elements, such as the skipped macroblocks (no direct MVs). This paper proposes a novel 11-dimensional feature set to detect MV-based video steganography based on the above observation. The proposed feature is extracted based on the skipped macroblocks by recompression calibration. Specifically, the feature consists of two components. The first is the probability distribution of motion vector prediction (MVP) difference, and the second is the probability distribution of partition state transfer. Extensive experiments on different conditions demonstrate that the proposed feature set achieves good detection accuracy, especially in lower embedding capacity. In addition, the loss of detection performance caused by recompression calibration using mismatched quantization parameters (QP) is within the acceptable range, so the proposed method can be used in practical scenarios.
\end{abstract}


\section{Introduction}
With the rapid development of information technology, steganology has become an emerging discipline in the field of information security, which mainly contains two opposing directions of steganography and steganalysis\cite{ref_Muralidharan_2022,ref_Bouzegza_2022}. Steganography achieves the goal of concealing the act of communication by hiding secret information in common media covers, including image, text, video and audio. The purpose of steganalysis is to detect the presence of secret information in common media through statistical analysis. Image steganography has been a hot research topic for the past two decades. However, with the popularity of video media on the Internet, research on video steganography has received more and more attention from scholars in recent years\cite{ref1,ref2}. Due to its complex coding rules, video has richer covers suitable for information embedding compared with image, text, and audio, mainly intra-frame prediction patterns\cite{ref3}, inter-frame prediction patterns\cite{ref4}, MVs\cite{ref5}, transformation coefficients\cite{ref6}, entropy coding coefficients\cite{ref7}, etc. MV-based steganography has a larger embedding capacity among these embedding domains due to the larger amount of MVs. Moreover, the MV-based steganography algorithm is closely connected with the video coding process. The steganographic perturbations on the MVs are usually handled automatically by the subsequent coding process, so the visual quality of the stego video is less affected. For these reasons, MV-based video steganography and steganalysis techniques have been the focus of researchers' attention.

The development process of MV-based video steganography can be divided into three stages. The first stage is the traditional steganography method, including the MV amplitude-based\cite{ref5} and phase-based\cite{ref8} steganography methods. The basic principle of this type of method is to select a specific MV by setting a rule and then directly modify the MV. The second stage mainly uses coding techniques\cite{ref9} to reduce the number of MV modifications and improve the algorithm's performance\cite{ref10}. Inspired by the framework of minimizing embedding distortion\cite{ref11} in image steganography, the third stage of MV-based video steganography is mainly based on the minimizing embedding distortion method, which is the mainstream framework of the whole steganography technology. The basic idea is to minimize the overall distortion by designing a distortion function for covers and then combining it with coding methods such as STC\cite{ref11}, which greatly improves the security of steganography algorithms. According to the design perspective of distortion function, the third stage of video steganography can be subdivided into complexity-based methods\cite{ref12}, local optimality-based methods\cite{ref13,ref14,ref15}, and multiple factors-based methods\cite{ref16,ref17,ref18}.

Video steganalysis is the adversary of video steganography, whose purpose is to detect whether the video media contains secret information by statistical analysis methods. Due to the complexity of video coding, MV-based video steganography leads to the perturbation of different types of coding parameters of the cover video stream\cite{refLi_2022}. MV-based video steganalysis can be divided into five categories from the different perspectives of feature extraction. The first category is the method based on the spatio-temporal statistical properties of MVs\cite{ref19,ref20} because there is a strong correlation between MVs similar to that between pixels and DCT coefficients in images. The second category is the method based on MV calibration\cite{ref21,ref22} because the stego MVs tend to recover to the original values after calibrating. The third category is based on the local optimality\cite{ref23,ref24,ref25} of MVs. Because the MV is a locally optimal output process in the sense of rate-distortion, the steganograhic operation will likely destroy this local optimality. The fourth category is the steganalysis algorithm designed based on the fact that the MVs of sub-blocks in a macroblock usually have non-consistency\cite{ref26}. And these methods have the best performance at present and can detect steganography methods based on inter-frame prediction patterns and MV domains simultaneously. The fifth category is the steganalysis method based on a convolutional neural network for the MV domain\cite{ref27}, which is at the initial research stage.

From the above steganalysis research status, all MV-based video steganalysis algorithms extract features directly on the MVs because MV-based steganography takes the MVs as the covers. It seems reasonable but ignores that video compression coding is a closely interconnected process. Modifications to the MVs cause perturbations to their own statistical properties and may lead to anomalies in the statistical characteristics of other coding parameters. For example, the MV Consistency steganalysis feature proposed by Zhai et al.\cite{ref26} can effectively detect inter-frame prediction pattern domain steganography even though the features are extracted from the MV domain. Take the H.264/AVC standard\cite{ref28,ref29} as an example, the P-frames' macroblocks are mainly P blocks(inter), I blocks(intra), and P-Skip macroblocks. Each P block contains a set of MVs (including horizontal and vertical components), which are used to point to the optimal reference block in a reference frame. I blocks are encoded using the intra-prediction mode, but their number in P frames is small. The P-Skip macroblocks do not directly contain MVs; their optimal reference block is determined by the MVP. The encoder calculates the P-Skip macroblock's MVP based on the MVs of its three encoded neighborhood blocks. MV-based steganography directly modifies the MVs of the P blocks, which leads to changes in the statistical features of the MVs. Therefore, feature extraction in the MV domain can detect MV-based steganography. Although P-Skip macroblocks do not have MVs directly used for steganographic embedding, their MVPs are determined by the MVs of their neighborhoods. If the encoded blocks in the neighborhood around the P-Skip macroblocks are modified, their MVPs may also be modified passively. Thus the best matching block corresponding to the P-Skip macroblocks will change from optimal to non-optimal. In this paper, we find that the MVP of the P-Skip macroblock and the distribution state of the P-Skip macroblocks will be significantly impacted by message embedding. However, all current steganalysis feature sets of MV-based steganography ignore the perturbations suffered by the P-Skip macroblocks. Based on this observation, this paper uses the MVP and partititon state distribution information of the P-Skip macroblocks to distinguish the cover videos from the stego videos. 

The main contributions of this paper can be summarized as follows.
\begin{enumerate}
	\item{For the first time, the coding information based on skipped macroblocks is used to construct MV domain video steganalysis feature, which enriches the methods to extract video steganalysis features.}
	\item{By the recompression calibration, an 11-dimensional feature set exploiting P-Skip macroblocks is proposed. The proposed feature has two components: a 5-dimensional sub-feature set based on MVP reversion and A 6-dimensional sub-feature set based on partition state transfer probability.}
	\item{The experimental results show that the proposed steganalysis feature set can effectively detect the state-of-the-art MV-based steganography algorithms, especially in low embedding capacity.}
\end{enumerate}

The rest of this paper is organized as follows. The second part gives the motivation for this paper. The basic principle of the skipped partition in video inter-frame coding is introduced, and the effect of MV-based steganography on skipped macroblocks is also analyzed. The third part presents the construction process of the steganalysis feature based on the skipped macroblock. The experimental results and analysis are given in the fourth part. Finally, the paper is concluded.

\section{Motivation}
Steganographic embedding inevitably causes perturbations to the original covers. The most important thing for steganalysis is finding relevant evidence to distinguish cover from stego videos. The motivation for our proposed steganalysis method based on skipped macroblocks is presented as follows.
\subsection{Skipped Macroblock in Inter-frame Video Coding}\label{subsection:A}
The current mainstream international video coding standards are H.264/AVC, H.265/HEVC\cite{ref30}, and H.266/VVC\cite{ref31}, etc. Although new coding standards have been advancing, there has been no revolutionary change. These compression standards use a hybrid coding framework, which usually contains techniques such as prediction, transformation, quantization, entropy coding, intra-frame prediction, inter-frame prediction, loop filtering, etc. H.264/AVC is still the most used compression standard, while the popularity of HEVC or VVC will take a long time. Especially in video steganography, most of the current research focuses on steganography and steganalysis based on the H.264/AVC standard. Therefore, in this paper, we still use the H.264/AVC compression standard for our research. However, it is worth noting that since the mainstream coding standards use the hybrid coding framework, this paper's proposed scheme can apply to new video coding standards with appropriate modifications. The hybrid coding framework uses the intra-frame prediction, inter-frame prediction, and entropy coding processes to reduce the spatial redundancy, temporal redundancy, and statistical redundancy in videos, respectively. Among them, temporal redundancy is the largest because natural video consists of consecutive frames, and adjacent frames usually contain the same content, especially in scenarios such as surveillance and conferences. Therefore the inter-frame predictive coding process has an important role in the coding framework.

In the inter-frame coding process of H.264/AVC, the frame $F$ is first divided into 16×16 pixel-sized non-overlapping macroblocks. Depending on the complexity of the macroblocks, the inter-frame encoded luminance blocks are divided into 16×16, 16×8, 8×16, and 8×8 (which can be further divided into 8×4, 4×8, and 4×4). For the block $ B_{m \times n} $ with the size of $ m \times n $ in a macroblock, the encoder uses the motion estimation (ME) algorithm to find the most suitable reference block $ T_{m \times n}^{mv(h,v)} $ in the reference frame based on the Lagrangian rate-distortion optimization model. $ mv(h,v) $ is the MV obtained by ME, which represents the position offset between the encoded block and the reference block, and contains horizontal component $ h $ and vertical component $ v $. The pixels' residual values between the current encoded block and the reference block are calculated as follows:
\begin{equation}
	\label{formula1}
	D_{m \times n}^{mv(h,v)} = B_{m \times n}^{} - T_{m \times n}^{mv(h,v)}.
\end{equation}
Then, these residual values are output as video compression stream by Discrete Cosine Transform (DCT), quantization, and entropy coding.

\begin{figure}[!t]
	\centering
	\subfloat[]{\includegraphics[width=1.5in]{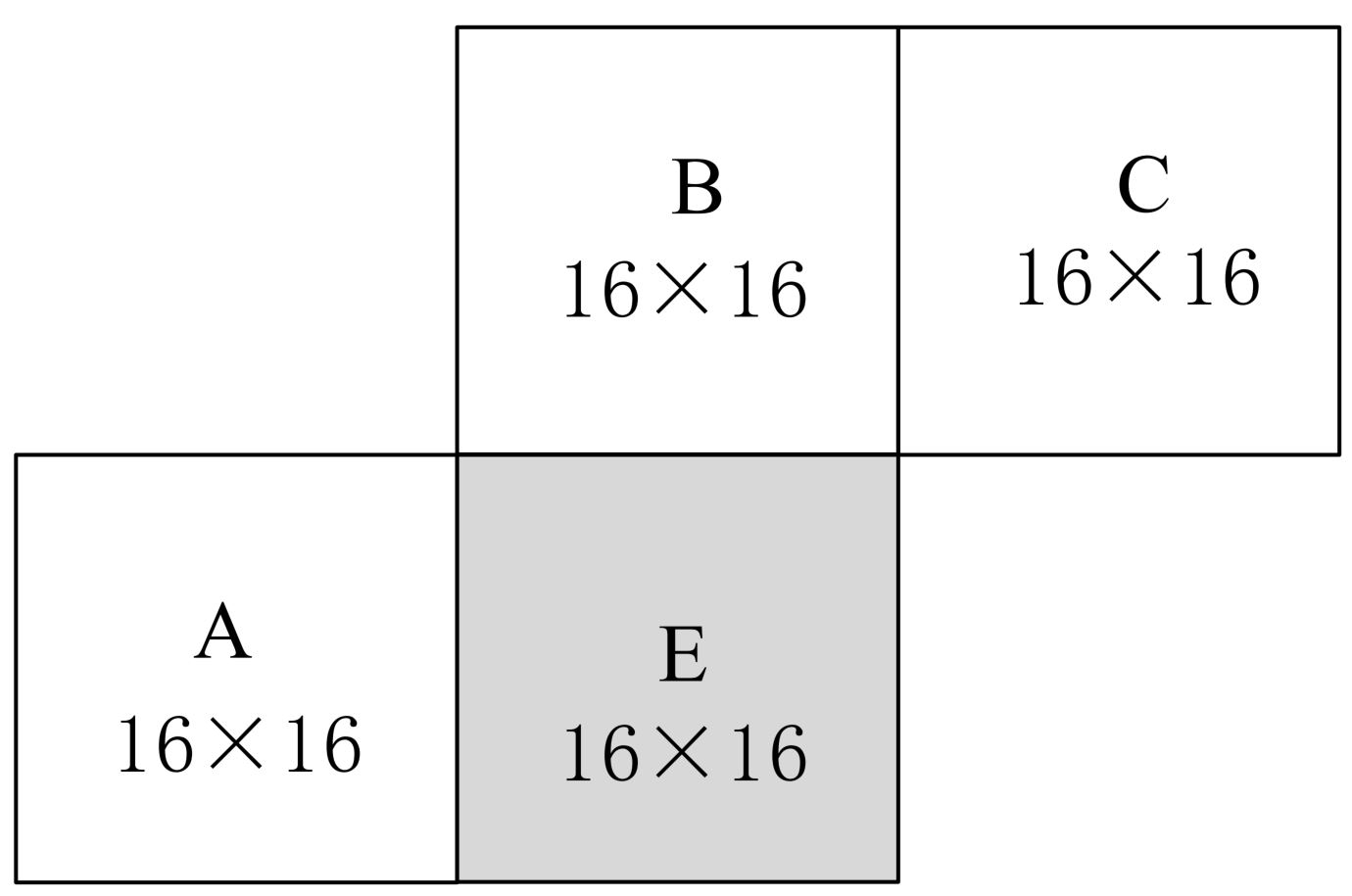}%
		\label{fig1(a)}}
	\hfil
	\subfloat[]{\includegraphics[width=1.5in]{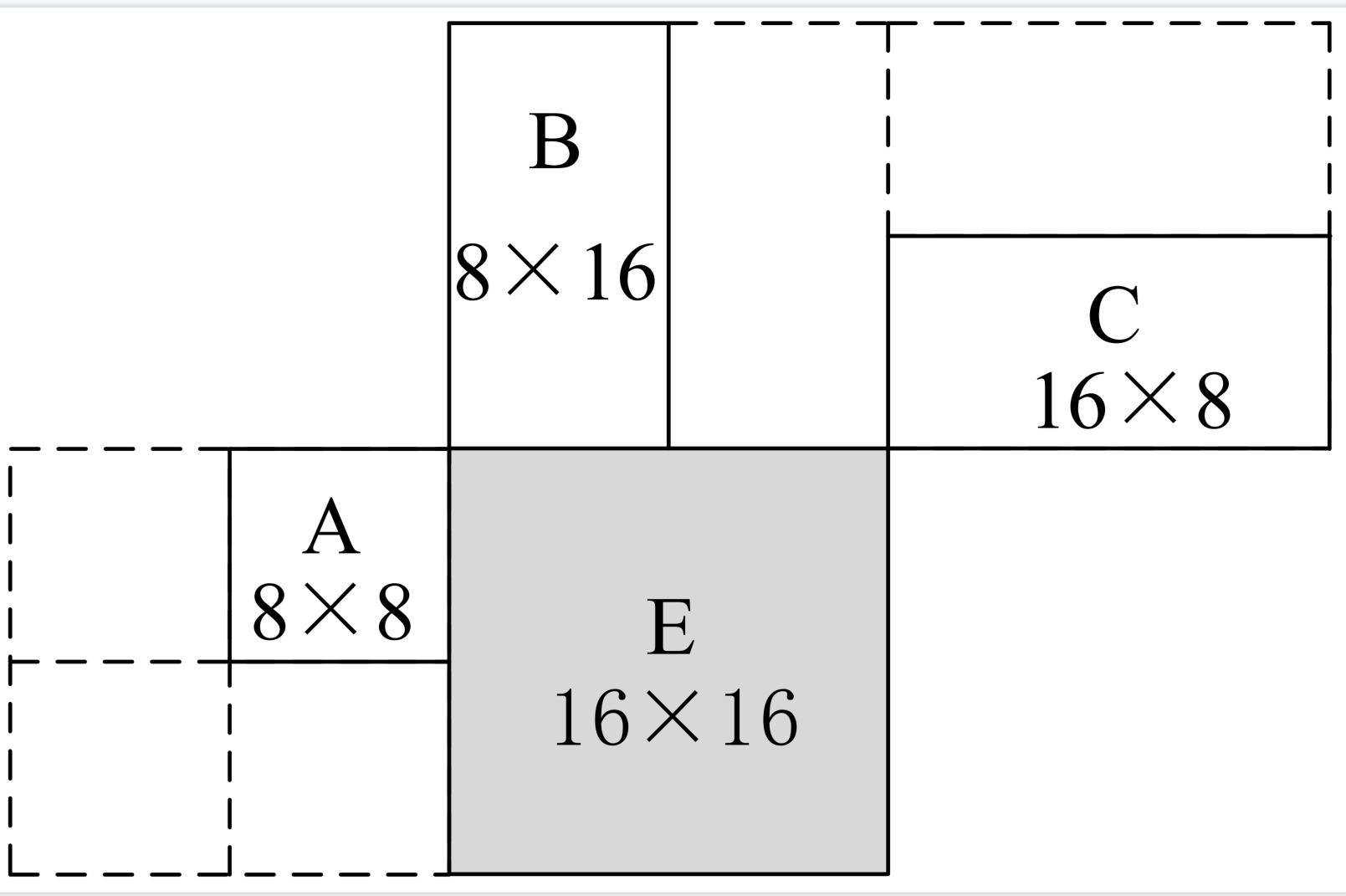}%
		\label{fig1(b)}}
	\caption{The MVP sketch map in H.264/AVC, numbers are the sizes of the blocks. (a) Current block E and adjacent blocks with the same size. (b) Current block E and adjacent blocks with different sizes.}
	\label{fig1}
\end{figure}

In addition, due to the strong correlation between adjacent encoding blocks, the MV of the current encoding block can be predicted based on the MVs of the encoded adjacent blocks. Fig.\ref{fig1} shows the MVP sketch map of H.264/AVC standard. E refers to the current encoding macroblock or sub-block. Blocks A, B, and C are three neighborhood blocks on the top, left, and top right of E, respectively. If there is more than one block on the left side of E, the uppermost one is A. If there is more than one block on the top, the leftmost one is B. Fig.\ref{fig1}(a) shows the case where all macroblock's partitions have the same size, and Fig.\ref{fig1}(b) shows the case where the partitions' sizes are not the same. The MVP of the encoding block E is determined by the median of the MVs of the encoded blocks A, B, and C \cite{ref28}. The motion vector difference (MVD) is output as a stream after entropy coding (Exponential-Golomb coding), whose value is the difference between the MV and the MVP as follows:
\begin{equation}
	\label{formula2}
	MVD = MV - MVP.
\end{equation}

Both pixel residual and MVD are output as a stream for a P block in a P-frame. However, there are special P-Skip macroblocks in P-frame. In H.264/AVC standard, a macroblock is encoded as a P-Skip macroblock when\cite{ref32}: 1) the optimal motion compensation macroblock's size is 16×16. 2) the reference frame must be the previous frame. 3) the MV and the MVP are the same. 4) the macroblock's pixel values are all zero after transformation. For the P-Skip macroblock, no other information about the macroblock is required to be transmitted at the encoding side except for some very few bits that identify the macroblock as a skipped macroblock. On the decoding side, the decoder can first get the MVP of this macroblock according to the prediction algorithm, and since MVD is 0, the MV obtained on the decoding side is equal to MVP. Then the decoder can recover the pixel value of this macroblock by finding the reconstructed pixel of the corresponding macroblock according to MV in the reference frame. In fact, in P-frames, there exist a large number of P-Skip macroblocks. We use an x264 encoder\cite{ref33} to compress the \textit{foreman.yuv} with a resolution of 352×288 in the standard test sequence\cite{ref34}. The percentage of P-Skip macroblocks in the first P-frame with QP of 25 and 35 is 33.8\% and 59.1\%, respectively. It can be seen that the proportion of P-Skip macroblocks in the normal video stream is very high, and the encoder can significantly improve the compression efficiency due to the small number of bits required for P-Skip macroblock coding.

\subsection{MV-based Video Steganography and Its Effect on Skipped Macroblocks}
In the MV-based video steganography, the MV $ mv(h,v) $ of a P block in a P-frame is used as the original cover. And the cover is modified by embedding algorithm $ E $ to:
\begin{equation}
	\label{formula3}
	mv(h',v') = E(mv(h,v)) = mv(h \pm \Delta h,v \pm \Delta v),
\end{equation}
where $ \Delta h $ and $ \Delta v $ are 0 or positive integers, representing the amplitude of modification, usually no more than 1. Obviously, after changing $ mv(h,v)$ to $ mv(h',v') $, the reference block $ T_{m,n}^{mv(h,v)} $ in formula(\ref{formula1}) will change to $ T_{m,n}^{mv(h',v')} $ , which will affect the whole subsequent encoding process. Therefore, the above embedding process will inevitably affect the original statistical properties of MVs. Current MV-based video steganalysis algorithms all extract features based on the statistical differences of MVs.
\begin{table}[b]
	\caption{\label{tbl1}Examples of MVP changes of P-Skip macroblock. Cover and Stego are the MVs before and after message embedding.}
	\centering 
	\renewcommand\arraystretch{1.3}
	\begin{tabular}{cccccc}
		\toprule
		&       			 &A                    &B                    &C                        &MVP of E\\
		\midrule
		\multirow{2}{*}{Case1}  &{Cover} 			 & {(1, 2)}            & {(1, 3)}            & {(2, 2)}                & {(1, 2)} 
		\\
		&{Stego} 				& {(1, 2)}           & {(\textbf{2}, 3)}   &  {(2, \textbf{3})}  & {(\textbf{2}, \textbf{3})} 
		\\
		\midrule 			
		\multirow{2}{*}{Case2}  &{Cover} 			 & { (14,6)}           & {(13,7)}            &  {(18,10)}            & {(14,7)} 
		\\
		& {Stego} 				& { (\textbf{13},6)} &{(13,\textbf{6})}    & {(18,10)}            & {(\textbf{13},\textbf{6})} 		
		\\
		\bottomrule
	\end{tabular} 
\end{table}
In addition to the direct effect on the MV of the P block, the steganography embedding will indirectly affect the MVP of the P-Skip macroblock. Taking Fig.\ref{fig1} (a) as an example, if the current encoding block E is a P-Skip macroblock, there is no direct MV for this block, and thus the steganography algorithm will not have a direct effect on this block. However, since the MVP of E is determined by the median of the MVs of blocks A, B, and C, if the MVs of blocks A, B, and C are modified during the embedding process, the MVP of E may also be perturbed. So the reference block of E will become non-optimal. Table \ref{tbl1} shows two examples of when the MVs of A, B, and C are perturbed by steganography. For case1, the optimal MVP of P-Skip macroblock E changes from (1, 2) to (2, 3). For case2, the optimal MVP changes from (14, 7) to (13, 6). The optimal matching blocks of these coding blocks have been deviated, thus providing the possibility for steganalysis.

To illustrate the steganography's effect on the P-Skip macroblocks, we embed the message to the \textit{foreman.264} standard sequence using an MV-based steganography algorithm named PCAMV(Principles on Cost Assignment for Motion Vector)\cite{ref17}. Table \ref{tbl2} shows the modified number of MVPs and their proportions in the first P-frame with different QPs and embedding capacities (bits per frame, bpf). From the data in the table, we can see that although the steganography operation does not directly modify the P-Skip macroblocks' MVPs, some of the MVPs are still modified. For example, the MVPs of an average of 23.6\% of the P-Skip macroblocks are perturbed at an embedding capacity of 200 bpf and QP of 25.

\begin{table}[!t]
	\caption{\label{tbl2} The number of perturbed MVPs of P-Skip macroblocks and their proportions after message embedding. The second column is the total number of P-Skip macroblocks in the first P-frame.}
	\centering 
	\renewcommand\arraystretch{1.3}
	\begin{tabular}{cccccccc}
		\toprule
		\multirow{2}{*}{QP}   &\multirow{2}{*}{Total number} & \multicolumn{6}{c}{Embedding Capacity (in bpf)}    \\
	    \cline{3-8} 
		            &       & \multicolumn{2}{c}{100}   & \multicolumn{2}{c}{150} & \multicolumn{2}{c}{200}     \\
		\midrule 		
		 25  		 & 89 		 &  9     &  10.1\%           &15     & 16.9\%    &21  &    23.6\%    \\ 		
		 35          & 230 		  	&  19     & 8.3\%            &23      &10.0\%    &45   &   19.6\%       \\
		\bottomrule                 	
	\end{tabular} 
\end{table}

\subsection{Recompression Calibration}
The idea of recompression calibration\cite{ref35} comes from JPEG image steganalysis, which means that the coding parameters of JPEG images can be revised to the original state after recompression. For the stego video, it is possible to recover the original MV through calibration, which can provide evidence for determining whether the video contains the secret message. Recompression, MV correlation, and MV coding optimality are the primary calibration methods used in video steganalysis. 
Cao et al.\cite{ref21} found that the MVs of calibrated stego video with the same parameters would exhibit a reversion to their original values. Therefore, they designed 15-dimensional feature of Motion Vector Reversion Based (MVRB) based on the difference between MVs and prediction errors before and after calibration.
Deng et al.\cite{ref36} proposed calibration-based steganalysis feature from the perspective of neighboring blocks, but there is a problem of poor applicability. To solve the problem of coding parameter mismatch, Wang et al.\cite{ref37} first collect various types of invariant coding parameters (e.g., size, frame rate, etc.), and then find the best ME algorithm by searching. This method has a performance improvement but has high computational complexity. 
In order to construct steganalysis feature applicable to various coding standards, Zhai et al.\cite{ref22} propose joint calibration feature with dimension 124 from three aspects: neighborhood optimality of MV, MV residual distribution, and MV calibration. 

From the above literature, existing calibration-based video steganalysis algorithms for the MV domain aim to recover the original MVs of P blocks but ignore that the MVPs of the P-Skip macroblocks can also be predicted by calibration. In this paper, we use the calibration method to recover various statistical properties of P-Skip macroblocks.

Based on the above analysis, the following conclusions can be drawn, which are the main motivations to design steganalysis feature from the perspective of skipped macroblocks.
\begin{enumerate}
	\item{Skipped macroblocks are common in normal video coding, and their number increases with the increase of QP.}
	\item{Although skipped macroblocks do not have MVs available for steganographic embedding, the steganographic operation may impact their MVPs, thus affecting the optimality of the skipped macroblocks.}
	\item{Recompression calibration is an effective method to recover coding parameters. We use recompression calibration to recover various statistical properties of Skip macroblocks for constructing steganalysis features.}
\end{enumerate}

\section{The Proposed Steganalysis Feature set Exploiting Skipped Macroblocks}
This section constructs the steganalysis feature set based on skipped macroblocks. Firstly, by the calibration of video recompression, a 5-dimensional sub-feature set based on MVP reversion of P-Skip macroblocks is constructed. Then, A 6-dimensional partition state transfer probability sub-feature set is designed by comparing P-Skip macroblocks before and after recompression calibration. Finally, we integrate these two sub-feature sets into an 11-dimensional feature set for MV-based steganalysis. Although skipped macroblocks also exist in B-frames, for simplicity, this paper mainly discusses the P-Skip coding block problem in P-frames.

\begin{figure*}[t]
	\centering
	\subfloat[]{\includegraphics[width=2.5in]{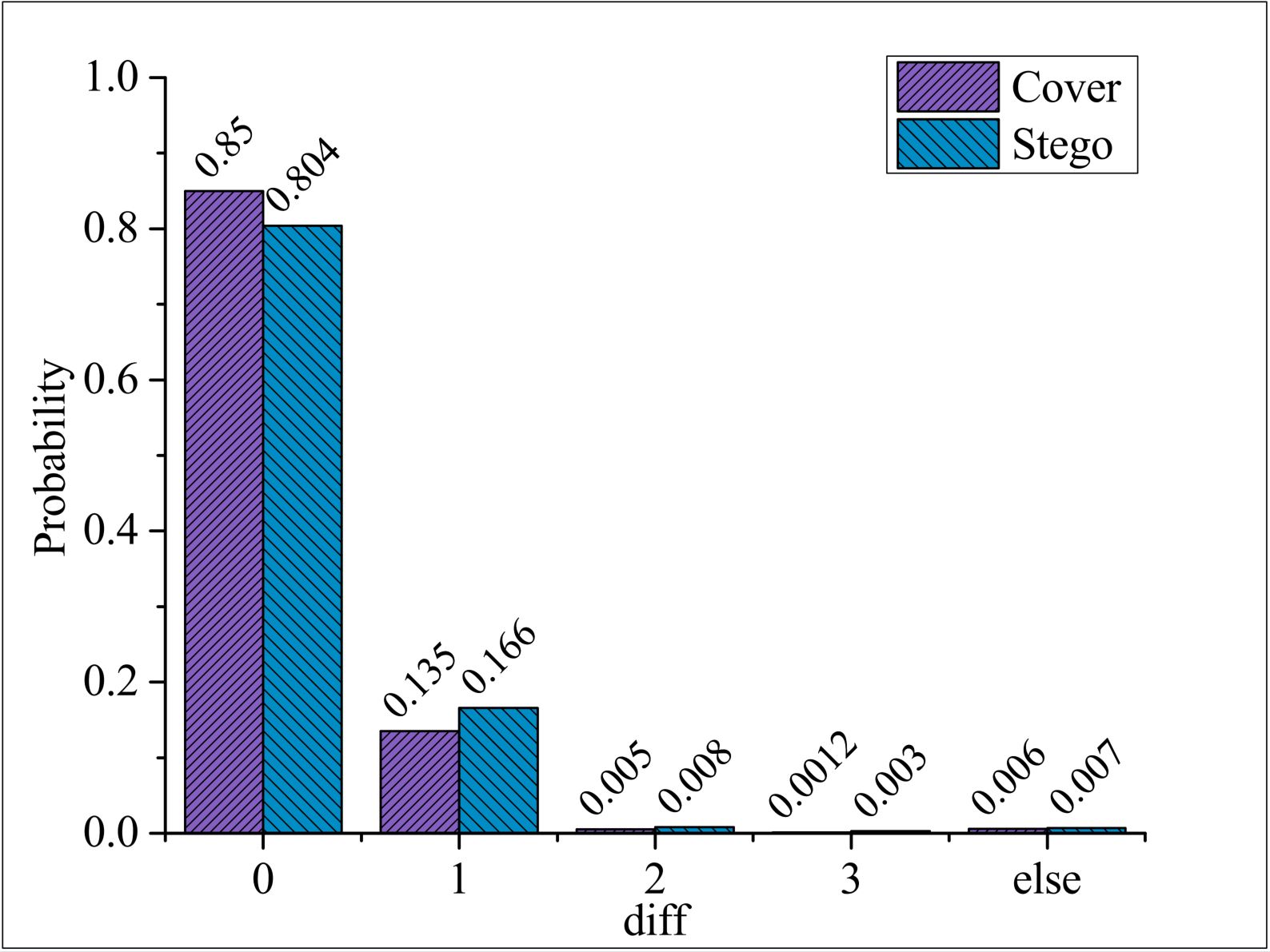}%
		\label{fig2(a)}}
	\hfil
	\subfloat[]{\includegraphics[width=2.5in]{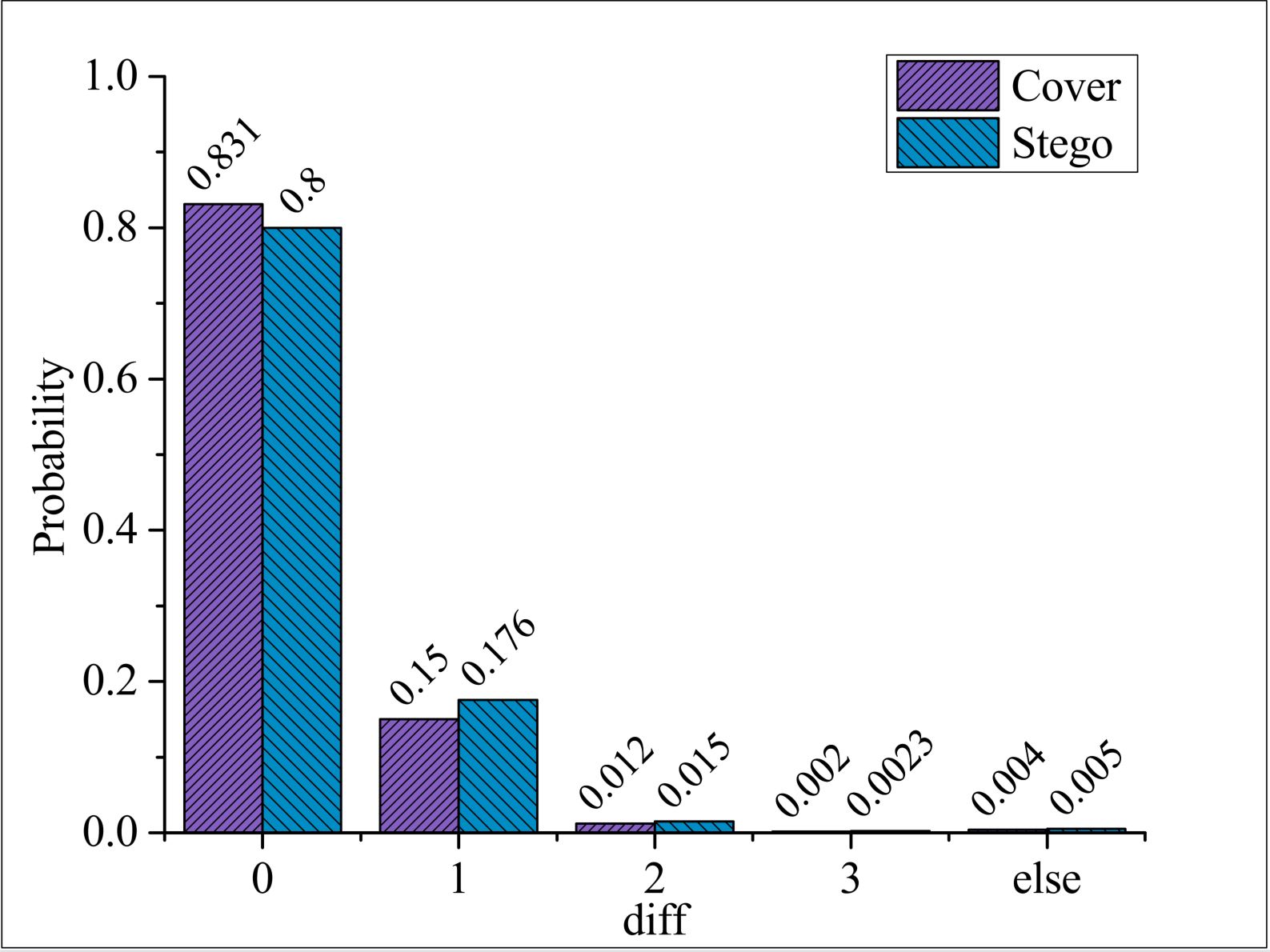}%
		\label{fig2(b)}}
	\caption{The MVP's difference distribution of P-Skip macroblocks by recompression calibration. (a) QP=25. (b) QP=35.}
	\label{fig2}
\end{figure*}
\subsection{The MVP's Revision Feature of P-Skip Macroblocks by Recompression Calibration}
\subsubsection{The effect of steganography on the MVPs of P-Skip macroblocks}
Section \textit{2.1} described that after message embedding on the MVs of the P blocks, a portion of the MVPs of the P-Skip macroblocks would also be perturbed. For a P-Skip encoding block in a P-frame, let its MVP before and after calibration be $ mvp(h,v) $ and $ mvp(h',v') $, respectively. We define a differential operator on the MVP of the P-Skip macroblock in formula \ref{formula4}, which describes the difference between MVP before and after calibration.
\begin{equation}
	\label{formula4}
	diff = |h - h'| + |v - v'|.
\end{equation}

Fig.\ref{fig2} shows the statistical distribution of \textit{diff} in the cover and stego video. Only those macroblocks' \textit{diff} that before and after calibration are all P-Skip macroblocks are counted. The stego video is embedded using the PCAMV algorithm \cite{ref17} with an embedding capacity of 100 bpf. The horizontal coordinate in Fig.\ref{fig2} is the value of \textit{diff}, and the vertical coordinate is the empirical probability of the occurrence in all P-Skip macroblocks. As can be seen from the figure, firstly, the difference of MVP in the cover video before and after calibration is mainly 0 (85\% and 83.1\% for QP of 25 and 35, respectively), while the number of P-Skip macroblocks with a difference exceeding 2 is very small. The result indicates that the recompression calibration operation accurately recreates the MVPs of the P-Skip macroblocks. Secondly, regardless of the QP with 25 or 35, there is a significant decrease in the probability of P-Skip macroblocks with a difference of zero in the stego video compared to the cover video. And there is a more significant increase in the probability of P-Skip macroblocks with a difference of one. Also, the probability of P-Skip macroblocks with a difference value greater than one increases to some extent. This is because the steganography mainly performs a plus or minus one operation on the MVs, resulting in a change in the MVP of the P-Skip macroblock in the stego video. The most direct effect is that the probability of blocks with a difference of zero decreases, while the probability of blocks with a difference of one increases. The above data illustrate that we can use the statistical features based on the difference of MVPs to distinguish cover videos from stego videos.

\subsubsection{Feature construction based on MVP reversion of P-Skip macroblocks}\label{lb3.2.2}

During feature extraction, firstly, the video sequence is calibrated by recompression. Then, the consecutive N P-frames are used as a feature extraction window. Suppose a total of \textit{n} P-Skip macroblocks are kept in the P-Skip partition before and after calibration, and these P-Skip macroblocks are denoted as $ \{ {B_i}\} _{i = 1}^n $. Based on the above discussion, we can design the first class of feature set based on MVP reversion, which is formally expressed as follows. 
\begin{equation}
	\label{formula5}
f_1^{}(k) = \Pr (diff{\rm{ = }}k) = \frac{{\sum\limits_{i = 1}^n {\delta(dif{f_{{B_i}}},k)} }}{n},\quad (k=0,1,2,3),
\end{equation}
\begin{equation}
	\label{formula6}
f_1^{}(4) = \Pr (diff >  = 4) = \frac{{\sum\limits_{i = 1}^n {\delta (dif{f_{{B_i}}},l)} }}{n},\quad (l >  = 4),
\end{equation}
where $ \delta (x,y){\rm{ = 1}} $  when \textit{x} is equal to \textit{y}, otherwise $ \delta (x,y){\rm{ = 0}} $. The above feature set represents the distribution of MVP's difference before and after calibration.

\subsection{The Partition State Transfer Probability Feature of P-Skip Macroblocks by Recompression Calibration}
\subsubsection{The effect of steganography on the partition state transfer of P-Skip macroblocks}
Recompression calibration can restore the state of the first compression with a considerable probability, but not perfect. There are some reasons. Firstly, video encoding is a lossy compression process, and recompression after decoding will cause some distortion.
Secondly, the parameters used in the second compression cannot be kept the same as those the first time, so there are mistakes between them. 
In the recompression calibration process, the coding block partitioned as P-Skip at the first time will maintain the P-Skip in the second recompression with a higher probability. Still, it may also become other partitions (such as P block of 16×16, 16×8, 8×8, etc.). However, from another point of view, the coded blocks not originally P-Skip partitioned may also transfer to P-Skip.
On the one hand, through the experiment, we found that for cover video and stego video, more than 95\% and 93\% of P-Skip macroblocks in the original compression will still keep the P-Skip partition in the second compression, respectively. 
And the statistical difference between the cover video and the stego video is insignificant. The result indicates that this feature can not provide a direct basis for steganalysis.

\begin{figure*}[!t]\begin{flushright}
		
	\end{flushright}
	\centering
	\subfloat[]{\includegraphics[width=2.5in]{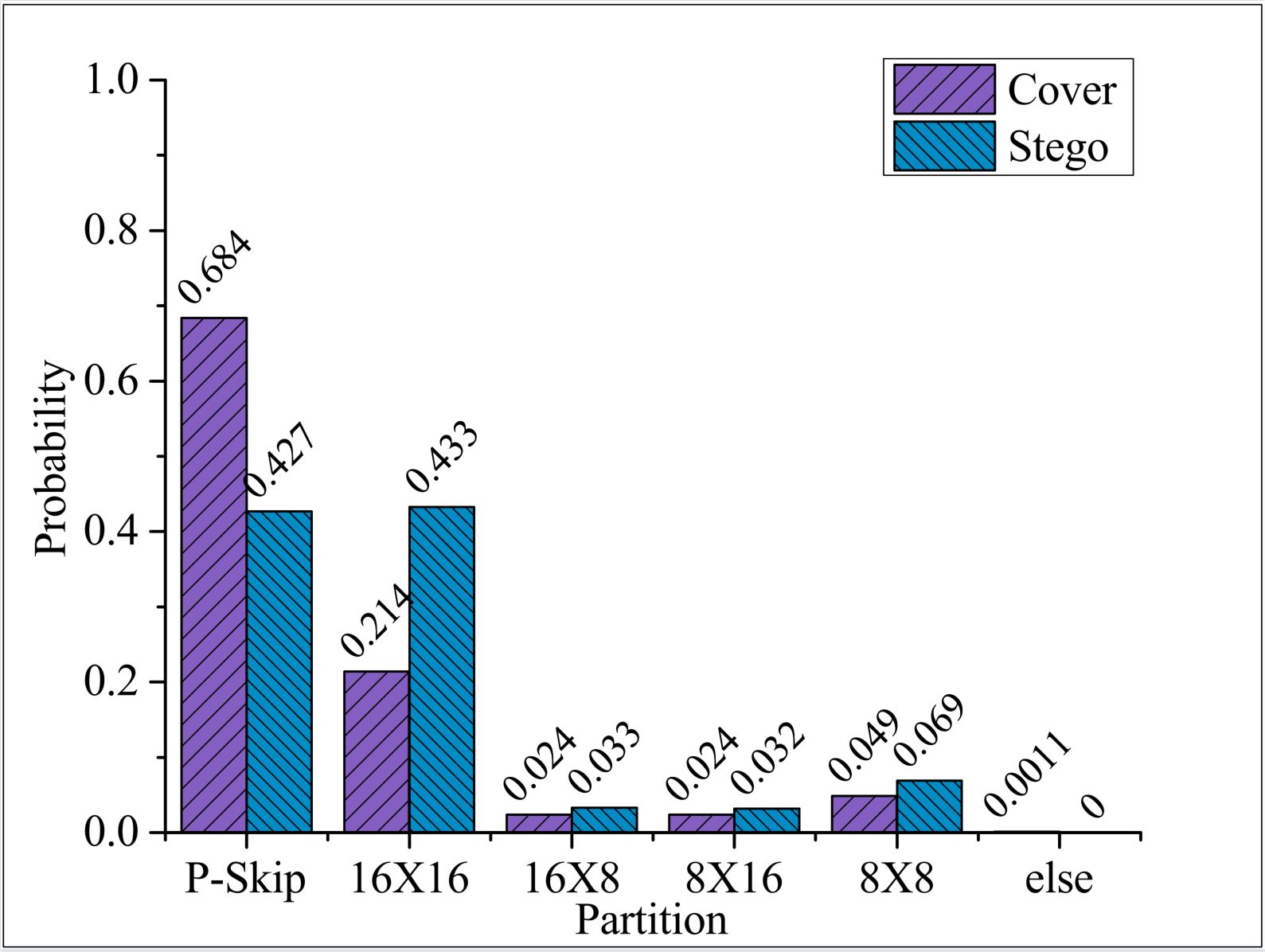}%
		\label{fig3(a)}}
	\hfil
	\subfloat[]{\includegraphics[width=2.5in]{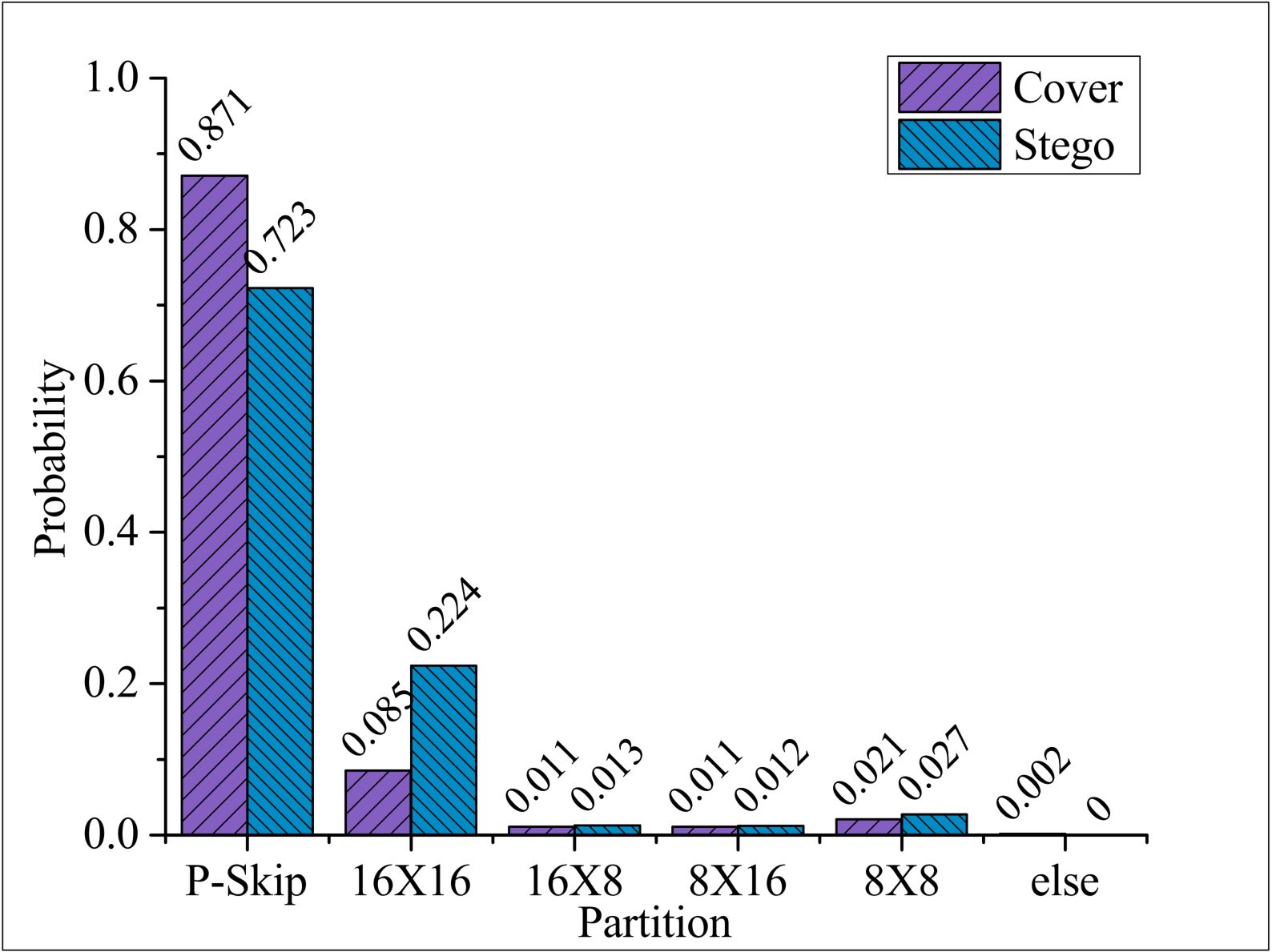}%
		\label{fig3(b)}}
	\caption{The original partition distribution (in the first compression) of blocks (whose partition is P-Skip in the second compression). (a) QP=25. (b) QP=35.}
	\label{fig3}
\end{figure*}

On the other hand, Fig.\ref{fig3} shows the original partition distribution (in the first compression) of blocks(whose partition is P-Skip in the second compression). 
For example, in Fig.\ref{fig3}(a) with the QP of 25, for the cover video, the partitions of the blocks at the first compression are mainly P-Skip and 16×16, and their proportions are 68.4\% and 21.4\%, respectively. For the stego video, the partitions of the blocks at the first compression are also mainly P-Skip and 16×16, but their proportions change significantly, 42.7\% and 43.3\%, respectively. 
The result indicates that fewer P-Skip macroblocks and more blocks with 16×16 partitions are encoded as P-Skip macroblocks after calibration in the stego video. 
The reason is the steganography embedding modifies the MVs of blocks, which leads to the change of the optimal partition method of blocks. The situation when the QP is 35 (shown in Fig.\ref{fig3}(b)) is similar to when QP is 25. Therefore, we can see that steganography indirectly affects the blocks' partition distribution in P-frames, which can be used as a steganalysis feature.

\subsubsection{Feature construction based on partition state transfer of P-Skip macroblocks}

The video sequence is calibrated by recompression, and then the consecutive N P-frames are used as a feature extraction window. Let there be a total of \textit{m} P-Skip macroblocks at the second compression, and denote these blocks as $ \{ {C_j}\} _{j = 1}^m $. 
Let the possible partitions of these blocks at the original compression be the set \textit{partition}:
\begin{equation}
		\label{formula7}
		partition=\{P-Skip, 16\times16, 16\times8, 8\times16, 8\times8, else\}.
\end{equation}
Based on the above discussion, we can design a feature set using the probability of partition state transfer of the P-Skip macroblocks, formally expressed as follows.
\begin{equation}
	\begin{split}
		\label{formula8}
		f_2^{}(k) = \Pr (partition(k - 5)) =\frac{{\sum\limits_{j = 1}^m {\phi ({C_j},partition(k - 5))} }}{m},\\\quad (k = 	5,6,7,8,9,10),
	\end{split}
\end{equation}
where $ \phi (C,y){\rm{ = 1}} $ when the $C$ block's partition is y, otherwise $ \phi (C,y){\rm{ = 0}} $. 
For example, when $k=5$, $ f_2^{}(5) = \Pr (P{\rm{-}}Skip) $. The feature $ f_2^{}(5) $ represents the probability of the original partition being P-Skip among $ m $ P-Skip macroblocks in the second recompression. Thus the feature set has six dimensions, corresponding to the probability that the blocks (whose partition is P-Skip in the second compression) is P-Skip, 16×16, 16×8, 8×16, 8×8, and else at the first compression, respectively.

\subsection{Feature Merging} 
\begin{figure*}[!t]
	\centering
	\includegraphics[width=0.9\linewidth]{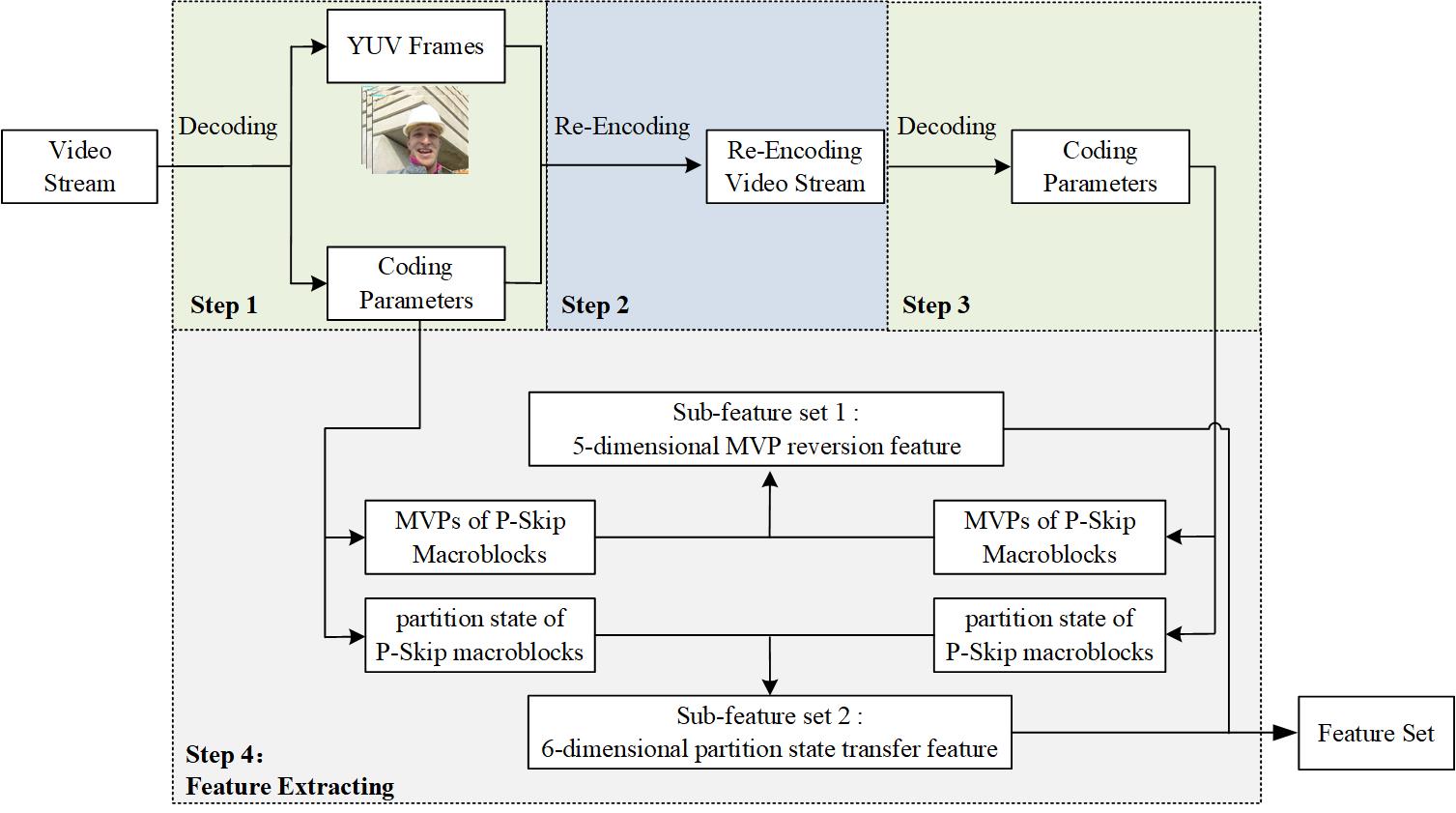}
	\caption{The flow chart of recompression calibration for feature set extraction.}
	\label{fig4}
\end{figure*}

We perform recompression calibration of the video to investigate the changes in various statistical features of P-Skip macroblocks before and after message embedding. The flow chart is shown in Fig.\ref{fig4}. In step 1, the compressed H.264/AVC stream is decompressed to obtain the decoded YUV file. The relevant coding parameters are extracted during the decoding process, including the frame number, resolution, group of picture(GOP) structure, QP, bits rate, macroblock partition, and MVs. In step 2, the spatial YUV file is encoded for the second time using the parameters from the first encoding, and the recompressed video stream is obtained. Then, in step 3, the recompressed video stream is decoded to obtain the macroblock's partition and MVs. Finally in step 4, the statistical information in the two streams is analyzed to extract the feature set that can effectively distinguish the cover from the stego video. The corresponding algorithm is shown in Algorithm 1.

According to the recompression calibration, the proposed 11-dimensional steganalysis feature set consists of two sub-feature sets: a 5-dimensional MVP reversion feature and a 6-dimensional partition state transfer feature based on P-Skip macroblocks. Since the feature is mainly implemented based on recompression calibration and P-Skip macroblocks, we name it SMCF (Skipped Macroblocks based Calibrated Feature).

\begin{algorithm}[!t]
	\caption{The extraction process of the proposed SMCF.}\label{alg:alg1}
	\begin{algorithmic}[1]
		\STATE \textbf{Input}: Original video stream \textbf{S}.
		\STATE \textbf{Output}: An 11-dimensional SMCF steganalysis feature set $ f $.
		\STATE \textbf{Steps:}
		\STATE The original video stream \textbf{S} is decoded to obtain spatial video sequence and parameter set $P_{\mathbf{S}}$;
		\STATE Recoding the spatial sequence by $P_{\mathbf{S}}$ to obtain the re-compressed video stream \textbf{S'};
		\STATE Take non-overlapping consecutive N P-frames in S and S' as one feature extraction unit. For each unit:
		\hspace{0.5cm}\FOR{$i=1, . . . , n$}
		\STATE For the P-Skip macroblock ${B_i}$ in \textbf{{S}}, find the corresponding block in \textbf{S'}, and calculate the feature $ f_{1} $ according to formula (\ref{formula5}),(\ref{formula6});
		\hspace{0.5cm}\ENDFOR
		\hspace{0.5cm}\FOR{$j=1, . . . , m$}
		\STATE For the P-Skip macroblock ${C_j}$ in \textbf{{S'}}, find the corresponding block in \textbf{S}, and calculate the feature $ f_{2} $ according to formula (\ref{formula8});
		\hspace{0.5cm}\ENDFOR
		\STATE Merge features $ f_{1} $ and $ f_{2} $ to obtain the 11-dimensional feature set $ f $.
	\end{algorithmic}
	\label{alg1}
\end{algorithm}

\section{Experiments and Analysis}
In this section, the experimental settings are first introduced. And then, in order to evaluate the performance of the proposed scheme, we present some experiments and analyses with different setups.
\subsection{Experiments Setup}
\subsubsection{Video database, H.264/AVC encoder, and decoder}
 As shown in Table \ref{tbl_DB}, we use two video databases for experiment. DB1: this database contains 34 well-known standard test video sequences\cite{ref34} with a CIF resolution(352×288), and each video sequence is cut into a fixed length by selecting its first 240 frames (so that we have a total of 8160 frames for experiments). DB2: this database contains 80 video sequences with different resolutions(from 416×240 to 2560×1600), which are downloaded from the internet, and each sequence is cut into a fixed length by selecting its first 100 frames (so the number of total frames is 8000). All the video sequences in DB1 and DB2 are stored in uncompressed file format, with YUV 4:2:0 color space.

\begin{table}[b]
	\caption{\label{tbl_DB} Databases of test video sequences.}
	\centering
	\renewcommand\arraystretch{1.3}
	\begin{tabular}{ccccc}
		\toprule
		Database  &\makecell{Number of \\Total Frames} & Resolution &\makecell{Number of \\Frames in \\Each Sequence} & \makecell{Number of \\Sequence} \\  
		\midrule 		
		DB1	      &8160	        &352 × 288	  &240   &	34  \\
		\midrule
		\multirow{5}{*}{DB2}   & \multirow{5}{*}{8000}    &2560×1600	&100	&10 \\
								&							&1920×1080	&100	&10 \\
								&							&1280×720	&100	&20 \\
								&							&832×480    &100	&20  \\
								&							&416×240    &100	&20 \\
		\bottomrule                 	
	\end{tabular} 

\end{table}

The H.264/AVC encoder in our experiments for video compression, recompression, and message embedding is the high-performance encoder x264 \cite{ref33} with the main profile. Unless otherwise specified, the GOP type of x264 is set to IPPPPP with the fixed size six and leaving other settings on the default parameters.
All P frames can be used for information embedding and feature extraction. The sub-block in P frames can be the variable size. The ME algorithm used in this paper is Hexagon-based Search (HEX) \cite{ref39} with a search range of 16 pixels, and the ME resolution is quarter-pixel. Also, three different QPs (15, 25, 35) are considered.

All the decoding process and steganalysis feature are implemented based on a well-known decoder FFmpeg \cite{ref40} and run on a desktop computer with a 1.8GHz Intel Core i7 CPU and 16 GB RAM. We use a sliding window with a length of 5 P-frames as a basic feature unit.

\subsubsection{Steganography methods}
To evaluate the detectability of video steganalysis in MV domain, three state-of-the-art typical MV-based steganography methods are used for message embedding. Yao et al.’s method \cite{ref12} (denoted as Tar1) is as the first one, which designed a distortion function based on the covariance matrix of MV residuals and inter-frame prediction errors. Zhang et al.’s method \cite{ref15} (denoted as Tar2) is as the second one, which designed the distortion function from the perspective of MV’s local optimality. And the third one is Li et al.’s method \cite{ref17} (denoted as Tar3) based on cost assignment for MVs. We also used the traditional Aly’s method \cite{ref5}(denoted as Tar4) to test the aplicability in B-frames. The embedding capacity bpnsmv (bits per non-skip motion vector) shall be set at 0.05,0.1,0.2,0.3,0.4 in our experiments. All the steganography methods are also implemented based on the x264 encoder.

\subsubsection{Competitor steganalysis methods}
We compare our proposed method with some state-of-the-art steganalytic methods, including the feature sets AoSO(Adding or Subtracting one) proposed by Wang et al.\cite{ref23}, the feature sets NPELO(Near-Perfect Estimation for Local Optimality) proposed by Zhang et al.\cite{ref24} from the perspective of local optimality, and also the multi-domain feature sets MVC(Motion Vector Consistency) proposed by Zhai et al.\cite{ref26}. 

\subsubsection{Training and classification}
We implement training and classification using a Gaussian-kernel SVM (support vector machine) \cite{ref41}, whose penalty factor $ c $ and kernel factor $ \gamma $ are determined by a five-fold cross-validation. And the detection performance is measured by the accuracy rate, which is defined as the ratio of correctly classified samples to the total samples. The final accuracy rate is averaged over ten random splits of the database. 60\% of the cover-stego video pairings are randomly chosen for training, and 40\% are used for testing in each iteration.

\subsection{Steganalysis Performances}
The detection accuracy against steganography is the most critical metric of steganalysis algorithms. Table \ref{tbl3} shows the detection accuracy of the proposed algorithm SMCF against three steganography algorithms with different embedding capacities and QPs using database DB1. 

\begin{table}[b]
	\caption{\label{tbl3} Detector accuracy(\%) of the proposed SMCF against three steganography methods with different embedding capacities and QPs using database DB1.}
	\centering
	\renewcommand\arraystretch{1.3}
	\begin{tabular}{cccccc}
		\toprule
		\multirow{2}{*}{Steganography Methods}   &\multirow{2}{*}{QP} & \multicolumn{4}{c}{Embedding Capacity (in bpnsmv)}  \\
		\cline{3-6} 
		&       & 0.05   & 0.1 & 0.2  &0.3  \\
		\midrule 		
		\multirow{3}{*}{Tar1 \cite{ref12}}  	
		&	15	&70.11	&73.46	&76.53	&77.90  \\ 		
		& 25	&76.23	&78.10	&81.56&	83.09    \\
		& 35	&72.15	&75.50	&78.67&	79.43    \\
		\midrule 	
		\multirow{3}{*}{Tar2 \cite{ref15}}  	
		&15  &70.98	&72.01	&73.32	&75.39  \\
		&	25&78.42	&80.59	&83.34	&84.21  \\
		&	35&75.25	&75.91	&76.75	&78.49  \\
		\midrule 	
		\multirow{3}{*}{Tar3 \cite{ref17}}  	
		&	15&67.20	&69.49	&70.54	&71.11\\
		&	25&73.01	&75.74	&79.70	&80.93\\
		&	35&68.22	&71.52	&73.06	&75.43\\
		\bottomrule                 	
	\end{tabular} 
\end{table}

\begin{figure*}[t]\begin{flushright}
	\end{flushright}
	\centering
	\subfloat[]{\includegraphics[width=2in]{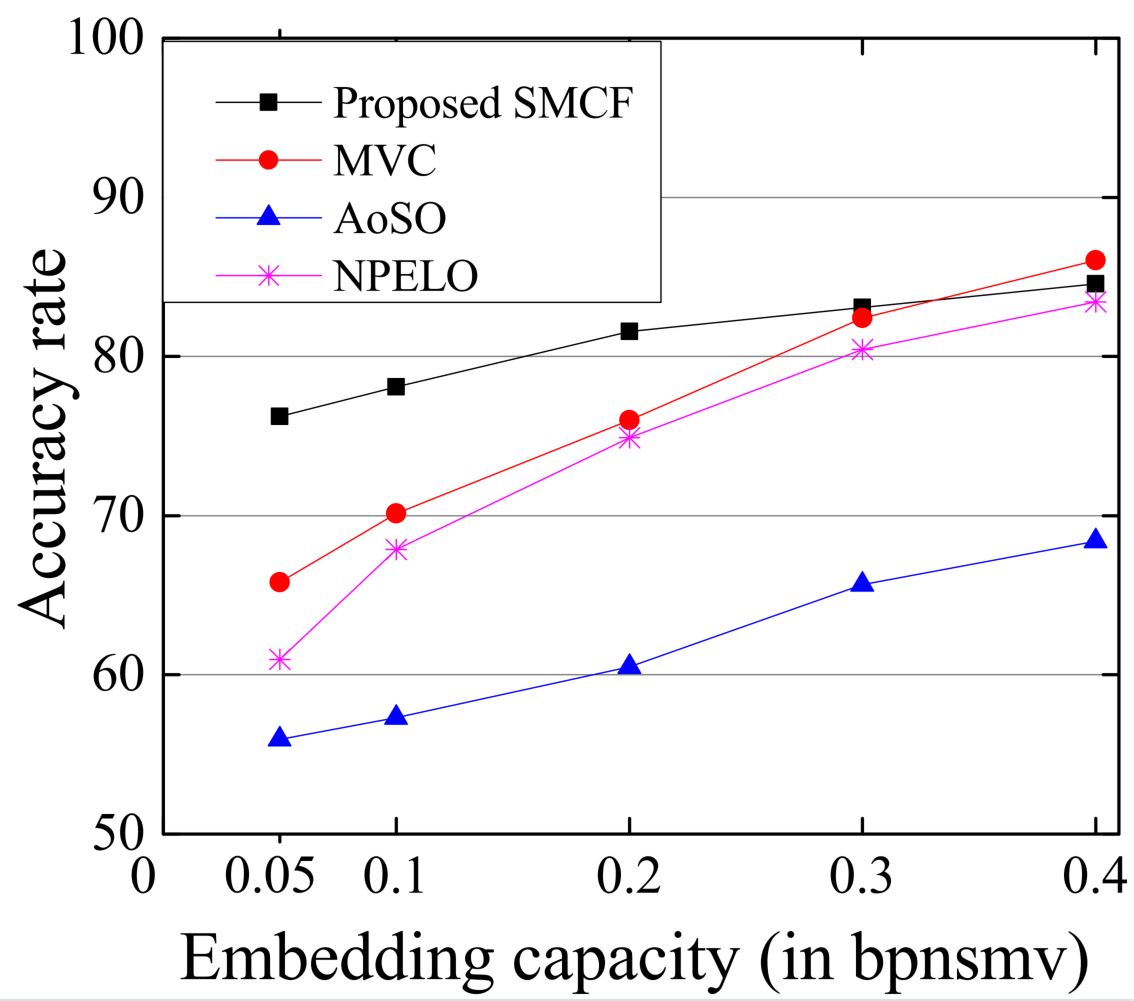}%
		\label{fig5(a)}}
	\hfil
	\subfloat[]{\includegraphics[width=2in]{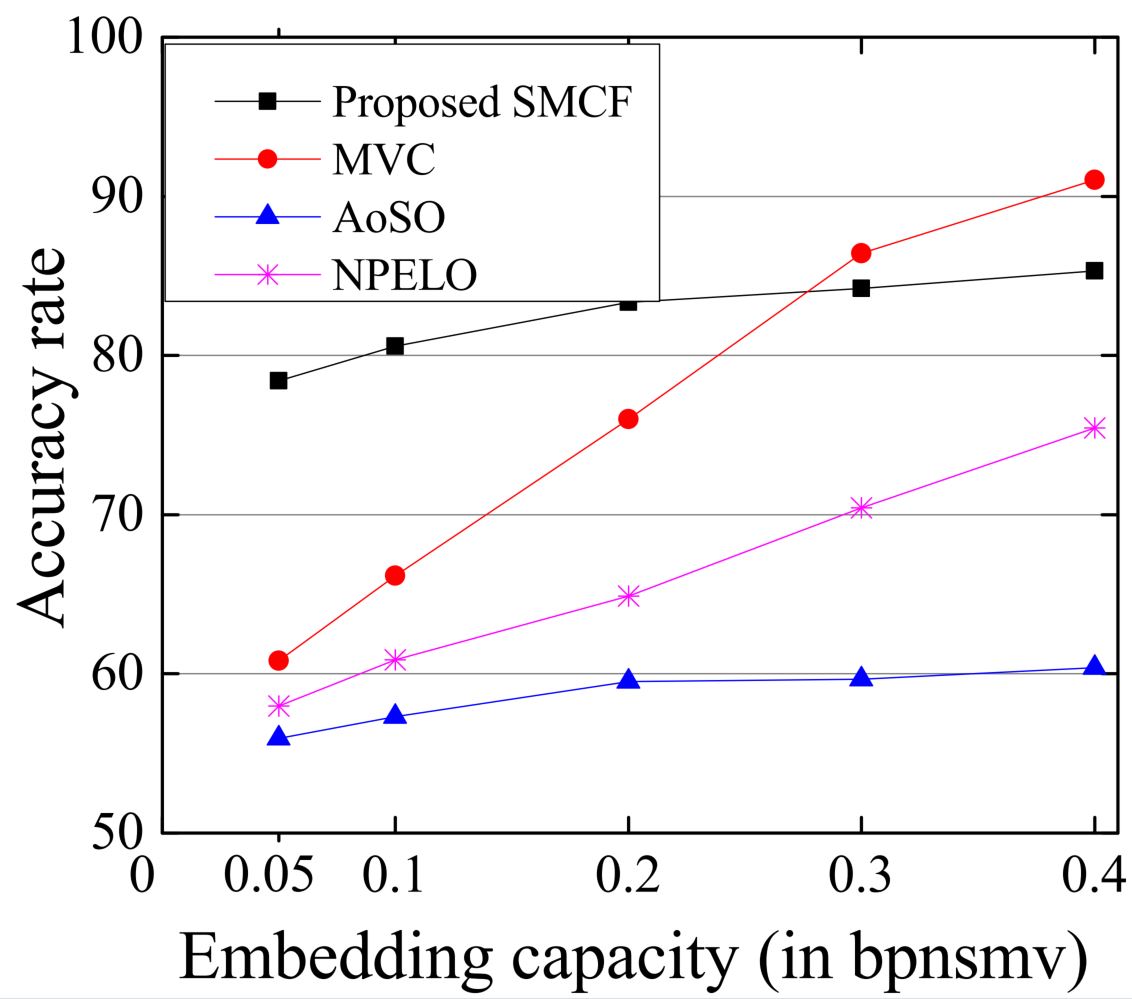}%
		\label{fig5(b)}}
	\hfil
	\subfloat[]{\includegraphics[width=2in]{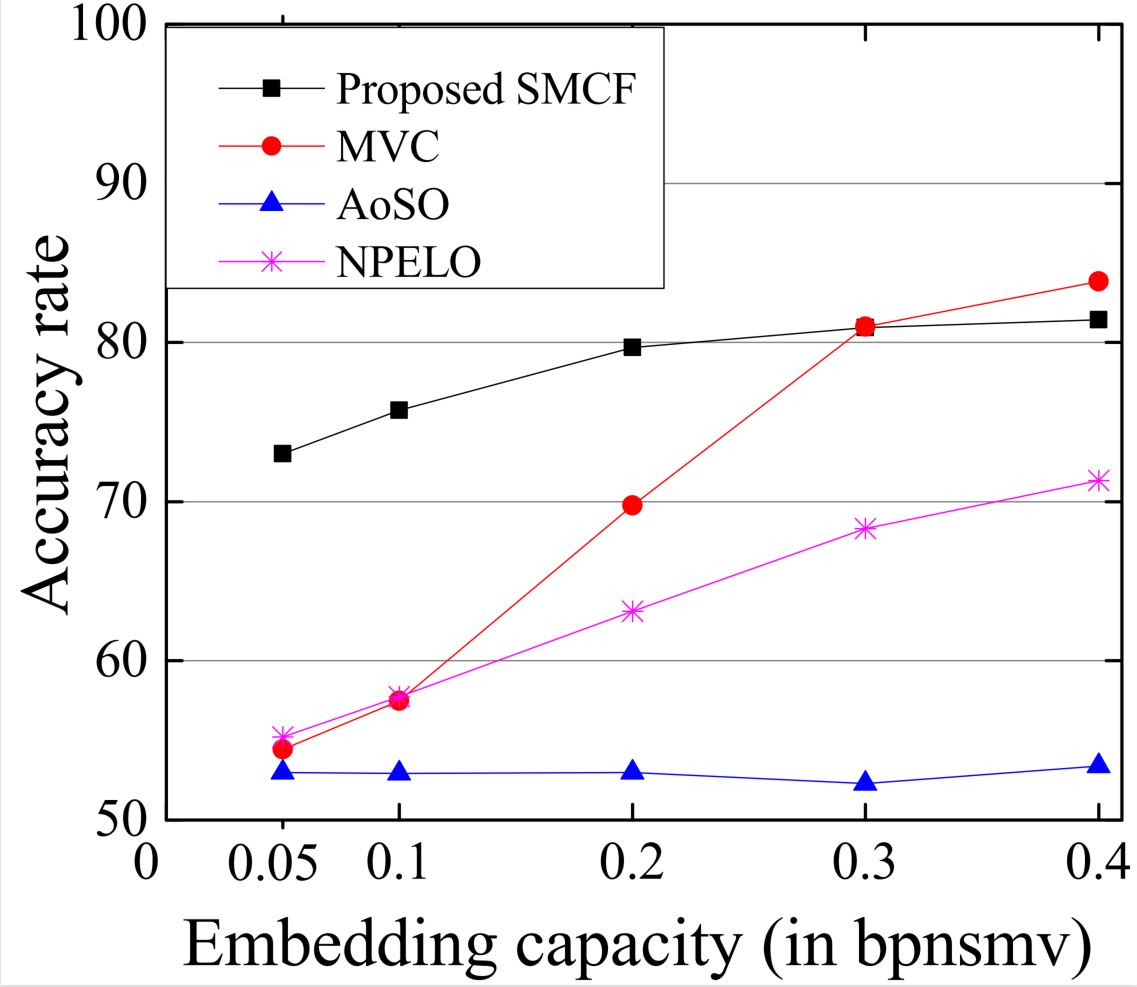}%
		\label{fig5(c)}}
	
	\subfloat[]{\includegraphics[width=2in]{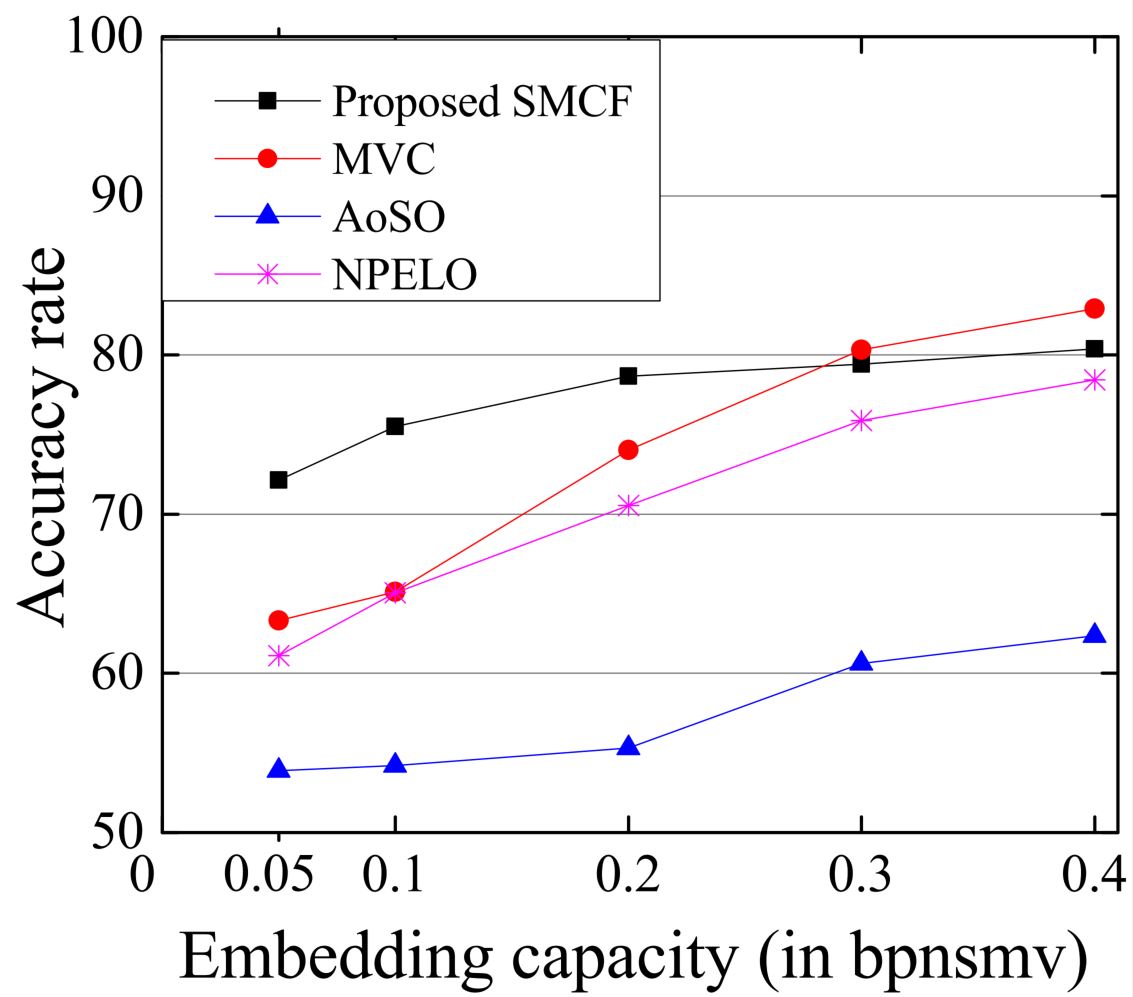}%
		\label{fig5(d)}}
	\hfil
	\subfloat[]{\includegraphics[width=2in]{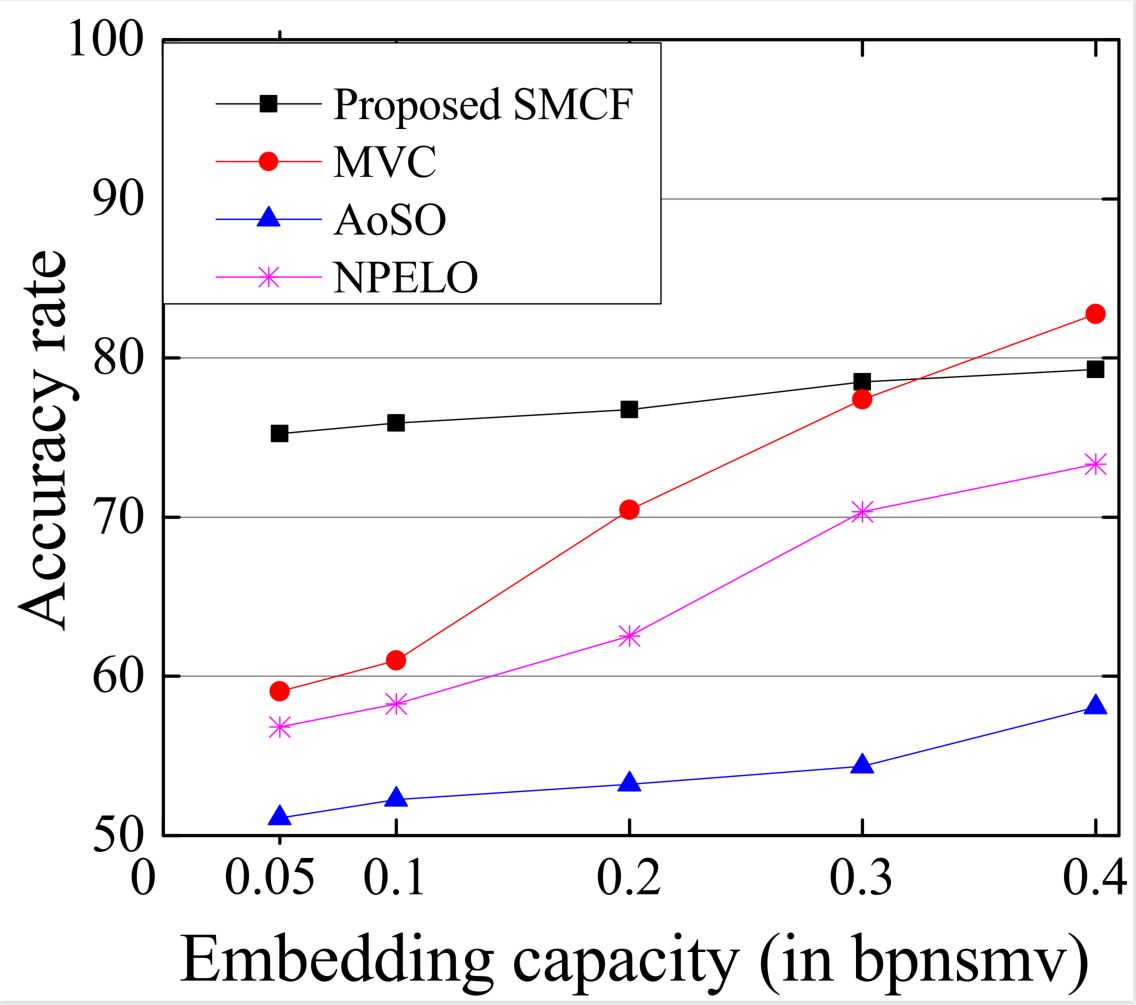}%
		\label{fig5(e)}}
	\hfil
	\subfloat[]{\includegraphics[width=2in]{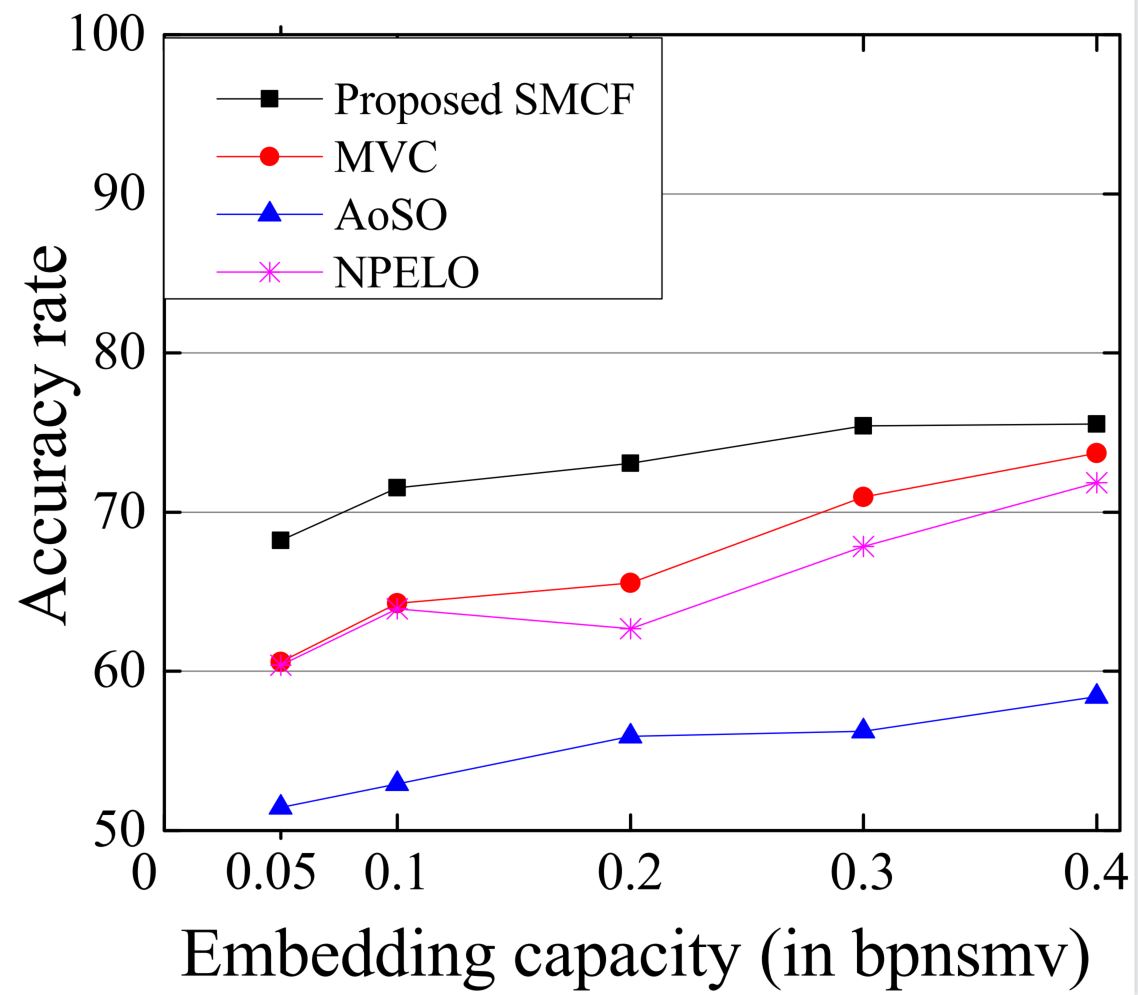}%
		\label{fig5(f)}}
	
	\caption{Accuracy rate of the proposed steganalysis against three steganography methods with different embedding capacities (in bpnsmv) and QPs using database DB1. (a) Tar1 \cite{ref12}, QP=25. (b) Tar2 \cite{ref15}, QP=25. (c) Tar3 \cite{ref17}, QP=25. (d) Tar1 \cite{ref12}, QP=35. (e) Tar2 \cite{ref15}, QP=35. (f) Tar3 \cite{ref17}, QP=35.}
	\label{fig5}
\end{figure*}

Firstly, the detection accuracy of the proposed algorithm for the three steganographic algorithms is 72.83\%, 74.88\%, and 69.48\% on average at the very low embedding capacity (0.05 bpnsmv). These results are a relatively high detection rate, showing the proposed algorithm's better performance at low embedding rates. As the embedding capacity increases, although the correct detection accuracy improves, the improvement is insignificant. The reason is the occurrence of P-Skip macroblocks in P-frames is aggregated, and once the MVP of one or more macroblocks is perturbed, the other macroblocks will also be perturbed with high probability. As the embedding capacity increases, the probability of the P-Skip macroblocks being perturbed does not increase significantly, despite the increase of the perturbed MVs. 
Secondly, the detection accuracy of the proposed SMCF does not differ significantly for different steganography algorithms. This is mainly because these steganography algorithms embed messages in P blocks. The proposed method do not extract feature directly on these blocks with MVs, but on the MVPs of the indirect P-Skip macroblocks, so they are not sensitive to different steganography algorithms. This also indicates that the feature proposed in this paper are more adaptable. 
Finally, for different QP, the detection accuracy of the proposed feature is higher at a QP of 25 than at QPs of 15 and 35. This is because at the smaller QP 15, the macroblocks are finer divided, and fewer macroblocks are partitioned as P-Skip, so fewer P-Skip macroblocks can be used for the proposed feature. While at larger QP 35, the video compression rate is larger. A large portion of macroblocks in P-frames are divided into P-Skip macroblocks, while the number of ordinary P blocks is small, and thus the number of MVs used for steganography is also small. At the same relative embedding capacity (bpnsmv), the MVs are subject to less steganography perturbation, so the steganography detection performance is reduced compared to that at a QP of 25.

Fig.\ref{fig5} shows the experimental comparison of the algorithm SMCF, AoSO, NPELO, and MVC against three different steganography methods with database DB1. Overall, the proposed scheme SMCF is significantly better than the AoSO and NPELO schemes under different conditions, which indicates that the method in this paper has better adaptability and detection performance. As the embedding capacity increases, the improvement rates of detection accuracy of the three steganalysis methods ( AoSO, NPELO, and MVC ) are all better than the proposed SMCF scheme. This is because all these three methods extract features directly in the MV domain. As the embedding capacity increases, the perturbation by the steganography operation to the MVs also increases, thus making it easier to be detected at large embedding capacities. 

The detection effect of the proposed scheme SMCF is better than that of the MVC scheme at low embedding capacities (bpnsmv of 0.05, 0.1, 0.2). Still, the detection effect of SMCF is equal to or lower than that of the MVC scheme at high embedding capacities (bpnsmv of 0.3, 0.4). Taking Fig.\ref{fig5}(b) as an example, the detection accuracy of SMCF is better than that of MVC at embedding capacity up to 0.3 bpnsmv; however, MVC outperforms SMCF at 0.3 bpnsmv and 0.4 bpnsmv. This is because MVC can better detect the perturbations in the Tar2\cite{ref15} algorithm for the consistency of the MVs within the macroblocks when the embedding capacity is larger. 

The detection accuracy of all four steganalysis methods is slightly lower at a QP of 35 than at a QP of 25, but the proposed SMCF feature are less reduced. This is because, at large QP (35) the macroblocks are mostly partitioned into P-Skip macroblocks, and the number of ordinary P blocks is small. Thus the number of MVs for steganography is also small. Therefore, the MVs are also subject to less perturbation at the same relative embedding capacity. So, the detection accuracy at high QP is reduced for features such as AoSO, NPELO, MVC, etc., which extract information directly in the MV domain. In contrast, the proposed scheme SMCF is less affected by the change of the QP because the feature are not extracted directly from the MVs.

\newpage

\subsection{Sub-feature Component Analysis}
The proposed SMCF feature set consists of two components: the sub-feature based on MVP reversion of P-Skip macroblocks (denoted as SMCF-part1), and the sub-feature based on partition state transfer of P-Skip macroblocks (denoted as SMCF-part2). In order to analyze the impact of two sub-features on the detection performance of the overall feature, we use Tar3 as the embedding method, DB1 as the database, and the experimental results are shown in Table \ref{tbl4}. From the results, the detection performance of SMCF is better than its subsets SMCF-part1 and SMCF-part2, which indicates that there is no obvious conflict between the two sub-features and the combination of the two sub-features can effectively improve the comprehensive detection ability. At a QP of 25, the detection ability of SMCF-part1 and SMCF-part2 is on average 15\% and 5\% lower than that of SMCF, indicating that SMCF-part2 occupies a more important role. It implies that the steganographic operation produces a greater impact on the partition state of the P-Skip macroblocks than on its MVP.

\begin{table}[t]
	\caption{\label{tbl4} Detector accuracy(\%) of sub-feature component against Tar3\cite{ref17} with different embedding capacities and QPs using database DB1.}
	\centering 
	\renewcommand\arraystretch{1.3}
	\begin{tabular}{cccccc}
		\toprule
		\multirow{2}{*}{Feature set}   &\multirow{2}{*}{QP} & \multicolumn{4}{c}{Embedding Capacity (in bpnsmv)}  \\
		\cline{3-6} 
		&       & 0.05   & 0.1 & 0.2  &0.3  \\
		\midrule 		
		\multirow{2}{*}{SMCF}  	
		& 25	&\textbf{73.01}	&\textbf{75.74}	&\textbf{79.70}	&\textbf{80.93} \\
		& 35	&\textbf{68.22}	&\textbf{71.52}	&\textbf{73.06}	&\textbf{75.43}  \\
		\midrule 	
		\multirow{2}{*}{SMCF-part1}  	
		&	25 &62.60	&65.62&	65.57&	69.49  \\
		&	35 &61.23	&63.91&	64.87&	66.14  \\
		\midrule 	
		\multirow{2}{*}{SMCF-part2}  	
		&	25 &70.03	&71.88	&74.26	&77.44\\
		&	35 &67.56	&69.25	&70.13	&71.03\\
		\bottomrule                 	
	\end{tabular} 
\end{table}

\begin{figure*}[t]\begin{flushright}
	\end{flushright}
	\centering
	\subfloat[]{\includegraphics[width=2in]{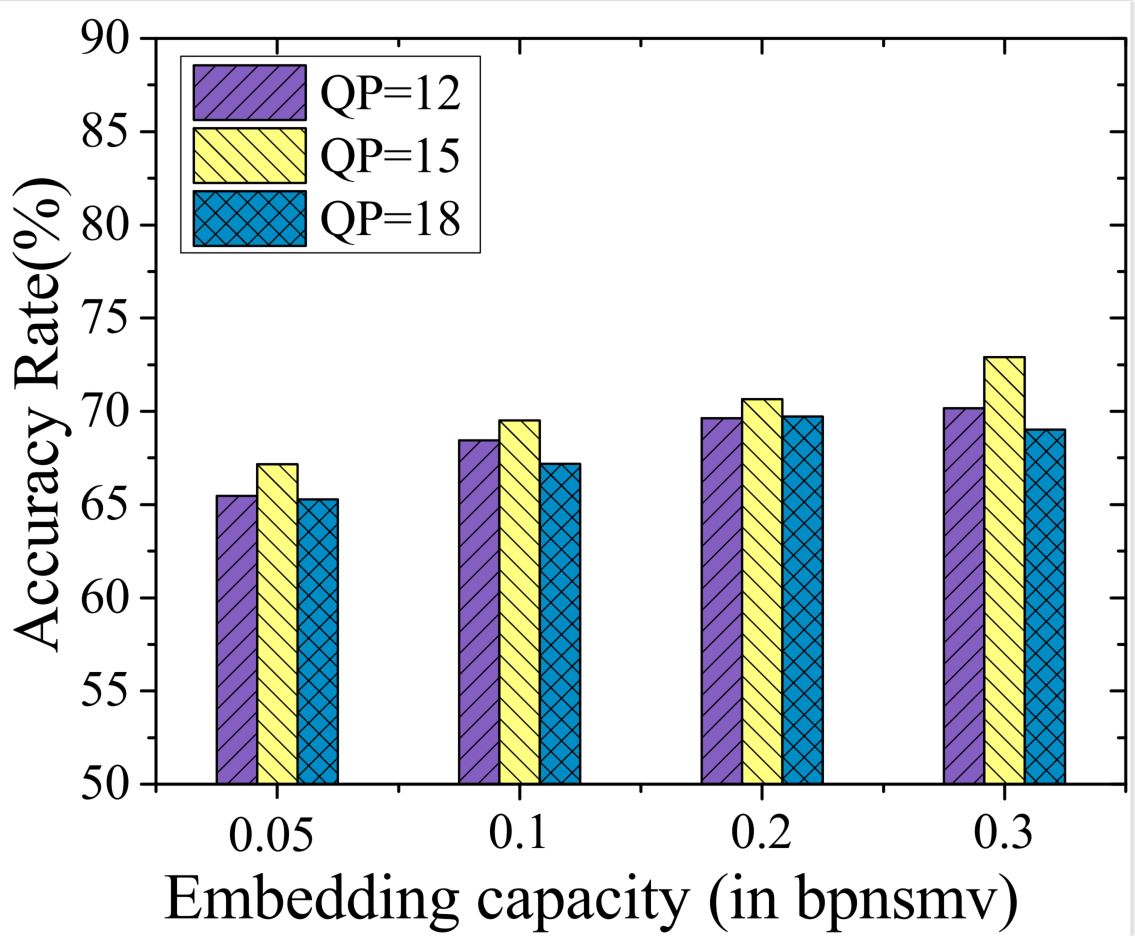}%
		\label{fig6(a)}}
	\hfil
	\subfloat[]{\includegraphics[width=2in]{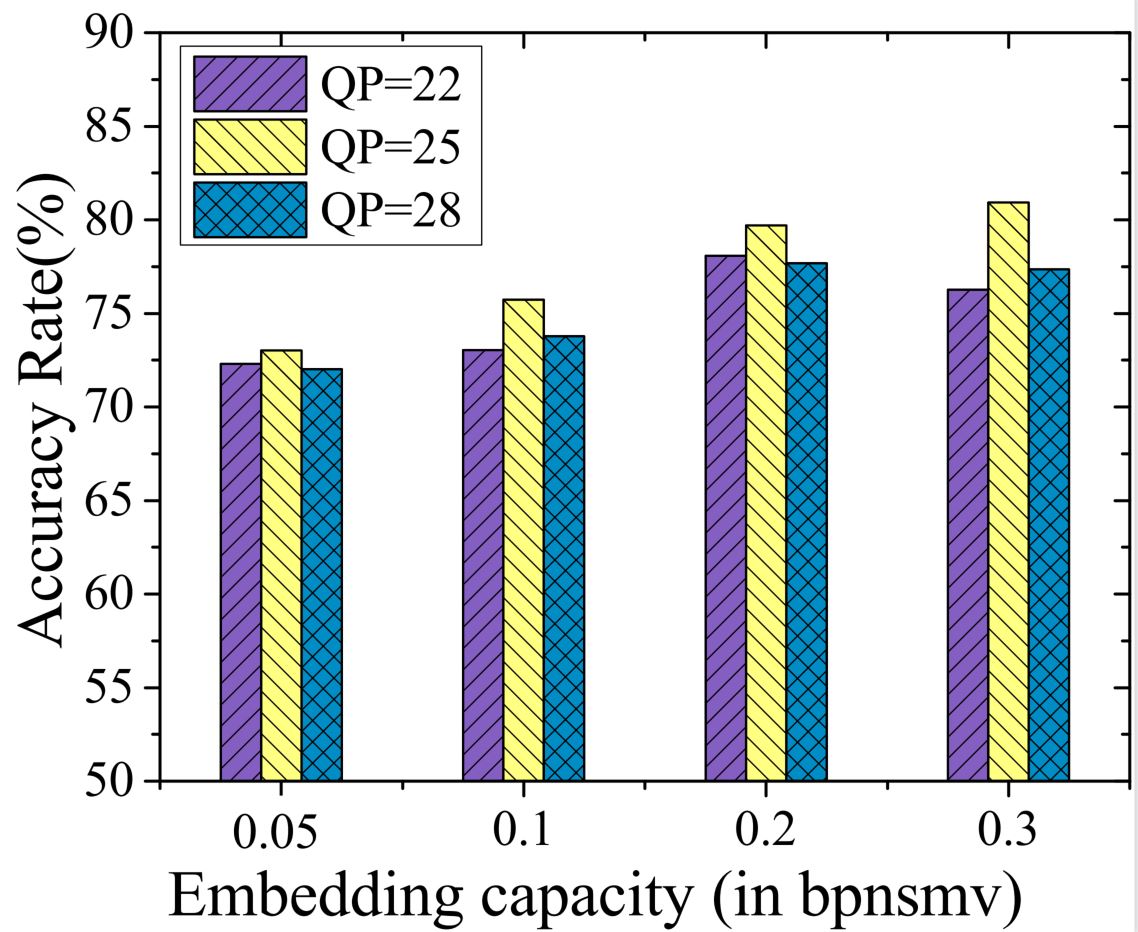}%
		\label{fig6(b)}}
	\hfil
	\subfloat[]{\includegraphics[width=2in]{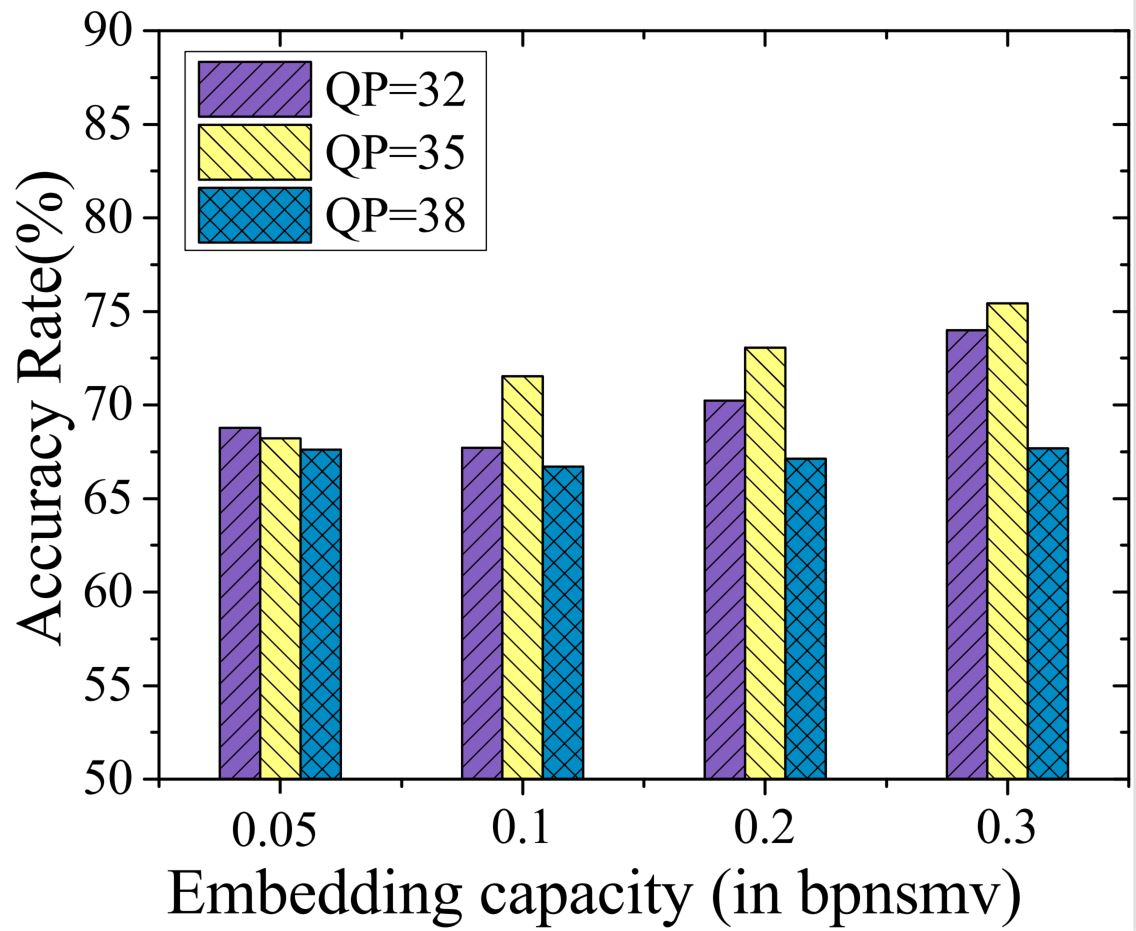}%
		\label{fig6(c)}}
	\caption{Accuracy rate of proposed method against Tar3[17] when using a different recompression QP to the original one. (a) Original QP=15. (b) Original QP=25. (c) Original QP=35.}
	\label{fig6}
\end{figure*}

The validity of SMCF-part2 in this experimental result also proves from the other side that it is reasonable to extract features from Skip macroblocks. Because the current main steganalysis methods, such as AoSO, NPELO and MVC, extract the MV information directly from non-skip macroblocks, which do not contain the partition state transfer features of the recompressed Skip blocks (SMCF-part2).

\subsection{Steganalysis Performance in the Case of Mismatch with QP Recompression}
The steganalysis feature designed in this paper is implemented based on recompression calibration. The traditional MV calibration-based video steganalysis \cite{ref21,ref37} does not apply to the encoding standard of variable size macroblocks such as H.264/AVC, because the sub-block partition will change greatly after recompression and cannot accurately locate the MV for steganalysis. 
The proposed scheme in this paper only considers the P-Skip macroblock partition and does not extract feature from sub-blocks, so it is less affected by the recoding parameters. In addition, the coding parameters of the video can be estimated by parameter searching\cite{ref37}, which enables getting the main parameters (size, frame rate, QP, etc.). Nevertheless, different compression parameters (mainly the QP) still impact the distribution of the P-Skip macroblocks.

Fig.\ref{fig6} shows the experimental results of recompression calibration using a QP different from the original one, the video database is DB1, and the embedding algorithm is Tar3. Taking Fig.\ref{fig6}(b) as an example, the QP used in the experiment for the original cover and the stego video is 25. Besides 25, the recompression calibration employs a QP of 22 and 28, which are close to 25 (the error in estimating the QP is not too large). As can be seen from the data in the figure, the highest detection performance of the proposed feature is achieved when the recompression QP is the same as the QP of the first compression (QP=25). Although the detection performance of the proposed feature decreases when the recompression QP (22, 28) is not the same as the QP of the first compression, the decrease is slight. This is because the proposed feature in this paper mainly consist of the MVP revision feature and the partition state transfer feature of the P-Skip macroblock. Although different QPs impact the number of P-Skip macroblocks (the bigger the QP, the larger the number of P-Skip macroblocks), they have a small impact on their distribution. Therefore, the proposed feature still maintain some applicability even if there is a certain degree of mismatch in the QP.

\subsection{The Complexity Analysis of the Proposed Feature}
This subsection compares the time required to extract four steganalysis features with different QPs. Table \ref{tbl5} shows the dimensionality and the average time needed to extract one CIF video sequence ( 352×288, 240 frames) for the four steganalysis features. 
As seen from the data in the table, the highest performance is achieved by the MVC feature because this feature mainly extracts MVs within macroblocks or sub-blocks and calculates their correlation alone. This is because both AoSO and NPELO need to traverse the neighborhood values of the MVs to calculate their local optimality. The highest time complexity is the SMCF, which is because the SMCF is implemented based on recompression calibration, and the running time is mainly concentrated in the two video decoding operations and one video recompression operation. 
In addition, the AoSO, NPELO, and MVC features have higher performance as the QP increases. This is because the larger the QP, the smaller the number of MVs in the video stream and the smaller the amount of data to be processed. However, the performance difference of SMCF feature at different QPs is not very obvious because it does not extract feature directly from the MVs of macroblocks or sub-blocks. In general, since the scheme in this paper is based on recompression to extract feature, its computational complexity is high.

\begin{table}[!t]
	\caption{\label{tbl5} Average computational time (in seconds) of feature extraction for a single CIF(352×288) sequence with 240 frames.}
	\centering 
	\renewcommand\arraystretch{1.3}
	\begin{tabular}{ccccc}
		\toprule
		\multirow{2}{*}{Steganalysis Methods}   &\multirow{2}{*}{Dimension} & \multicolumn{3}{c}{QP}  \\
	\cline{3-5} 
	&	&15	&25	&35\\
	\midrule 
	AoSO	&18	&3.93	&2.55	&1.82\\
	NPELO	&36	&3.21	&2.63	&1.75\\
	MVC	&12	&1.28	&0.87	&0.55\\
	Proposed SMCF	&11	&4.32	&4.21	&4.10\\		
	\bottomrule                 	
\end{tabular} 
\end{table}

\subsection{Applicability of different video databases}
For the purpose of testing the applicability of the proposed algorithm for videos with different resolutions and sources, this experiment is conducted using the database DB2. Table \ref{tbl7} shows the detection results of AoSO, NPELO, MVC, and the proposed SMCF for the steganographic algorithm Tar3 \cite{ref17} with different QPs and embedding capacities. From the experimental results, the detection performance of the proposed algorithm still outperforms the other three competing methods overall, proving that the proposed scheme has good applicability on different video databases. By comparing the data in Table \ref{tbl7} with the data in Fig.\ref{fig5}(c) and (f), the detection performance of all four steganalysis features on two databases, DB1 and DB2, does not differ significantly, which indicates that the size of the video resolution a small impact on the performance of steganalysis. The reason is that when steganography is performed using the relative embedding capacity bpnsmv, although the distribution of motion vectors is sparser for large-resolution videos, the number of their modified absolute motion vectors is higher, so the steganalysis features are still able to capture these steganographic perturbations.

\begin{table}[t]
	\caption{\label{tbl7} Detector accuracy(\%) against Tar3\cite{ref17} using DB2 with different embedding capacities and QPs.}
	\centering 
	\renewcommand\arraystretch{1.3}
	\begin{tabular}{cccccc}
		\toprule
		\multirow{2}{*}{Steganalysis Methods}   &\multirow{2}{*}{QP} & \multicolumn{4}{c}{Embedding Capacity (in bpnsmv)}  \\
		\cline{3-6} 
		&       & 0.05   & 0.1 & 0.2  &0.3  \\
		\midrule 		
		\multirow{2}{*}{AoSO}  	
		&25 &50.06	&54.41	&52.35	&52.94\\
		&35	&55.22	&50.44	&52.87	&53.68\\
		\midrule	
		\multirow{2}{*}{NPELO}  	
		&25	&57.50	&58.75	&65.25	&71.24\\
		&35	&60.37	&61.54	&65.88	&71.99\\
		\midrule
		\multirow{2}{*}{MVC}  	
		&25	&59.81	&71.48	&74.51	&81.67\\
		&35	&62.22	&63.06	&64.17	&70.28\\
		\midrule
		\multirow{2}{*}{Proposed SMCF}  	
		&25	&75.15	&80.07	&81.71	&82.51\\
		&35	&70.96	&71.43	&72.44	&76.62\\
		
		\bottomrule                 	
	\end{tabular} 
\end{table}

\subsection{Applicability to B-frames}
To test the applicability of the proposed algorithm in B-frames, this experiment verifies the detection performance of the proposed algorithm on the DB1 database. The GOP type of x264 is set to IPBBBPBBBIPBBB..., i.e., 9-frame GOP with One I-frames, two P-frames, and three B-frames in one GOP. As the steganography methods Tar1, Tar2, and Tar3 do not present the detail of how to embed messages in B-frames, we use Tar4 as the embedding method. Since the B-Skip block in B-frames has two MVPs, the results of Eqs. (5) and (6) are averaged over the two lists when calculating the features. The data in Table \ref{tbl8} compare the performance of the steganalysis with different GOP structures. The experimental data show that the frame type has little effect on the performance of the proposed scheme, so it can be concluded that the proposed method can be applied to B-frames.

\begin{table}[t]
	\caption{\label{tbl8} Detector accuracy(\%) against Tar4 \cite{ref5} using database DB1 with different embedding capacities and QPs}
	\centering 
	\renewcommand\arraystretch{1.3}
	\begin{tabular}{cccccc}
		\toprule
		\multirow{2}{*}{GOP}   &\multirow{2}{*}{QP} & \multicolumn{4}{c}{Embedding Capacity (in bpnsmv)}  \\
		\cline{3-6} 
		&       & 0.05   & 0.1 & 0.2  &0.3  \\
		\midrule 		
		\multirow{2}{*}{IPPPPP}  	
	   &25	&89.08	&92.06	&93.14	&95.01\\
	   &35	&90.54	&91.68	&93.37	&92.22\\
		\midrule	
		\multirow{2}{*}{IPBBBPBBB}  	
    	&25	&88.44	&90.78	&94.31	&94.53\\
	    &35	&91.15	&90.91	&92.04	&93.11\\
		
		\bottomrule                 	
	\end{tabular} 
\end{table}

\section{Conclusion}
This paper analyzes the skipped macroblocks in the inter-frame coding using recompression calibration. Firstly, the MVP of the skipped macroblock tends to return to the original value after recompression calibration, and the steganography operation will perturb this tendency. Secondly, the steganography operation also perturbed the distribution of partition state transfer probability of the skipped macroblock before and after recompression calibration. Based on the above analysis, an 11-dimensional MV-domain video steganalysis method based on skipped macroblocks is proposed in this paper. The experimental results show that the proposed feature have a high correct detection accuracy for state-of-the-art MV-based steganography algorithms, especially in low embedding capacities. In addition, since the proposed scheme only considers P-Skip macroblocks and does not extract feature from sub-blocks, it is less affected by the recoding parameters. It remains effective in the case of QP mismatch. However, the computational complexity is high since the proposed feature are implemented based on recompression calibration (decoding and recompression operations are required).

The inter-frame encoding frame is mainly composed of P blocks and skipped macroblocks. Due to the limitation of space, this paper only extracts the feature from the P-Skip macroblocks without considering the perturbation of the MV of the P blocks themself. Therefore, to further improve the detection performance and applicability of the steganalysis algorithm, we will next study how to extract steganalysis feature from both the skipped macroblocks and the P blocks.


\newpage

\vfill

\end{document}